\documentclass[11pt]{article}

\usepackage[letterpaper,margin=1in]{geometry}
\usepackage[T1]{fontenc}
\usepackage{lmodern}
\usepackage{microtype}
\usepackage{setspace}

\usepackage{amsmath}
\usepackage{amssymb}
\usepackage{booktabs}
\usepackage{array}
\newcolumntype{P}[1]{>{\raggedright\arraybackslash}p{#1}}  
\usepackage{graphicx}
\usepackage[font=small,labelsep=colon]{caption}
\usepackage{xcolor}
\usepackage{tcolorbox}
\tcbuselibrary{skins,breakable}
\usepackage[round,sort]{natbib}
\usepackage{parskip}  
\usepackage{xurl}
\usepackage[unicode,psdextra]{hyperref}
\hypersetup{
  colorlinks=true, linkcolor=black, citecolor=blue!50!black,
  urlcolor=blue!50!black, bookmarksnumbered=true,
  pdftitle={The Noise Is the Signal: Correlated Sampling Error Is
    Rank-Informative for Proxy Metric Selection},
  pdfauthor={Sandro Provenzano},
  pdfkeywords={proxy metrics, surrogate metrics, A/B testing, meta-analysis,
    sampling covariance, experimentation platforms, metric selection,
    ranking loss}}

\makeatletter
\renewcommand\section{\@startsection{section}{1}{\z@}%
  {-3.5ex \@plus -1ex \@minus -.2ex}{2.3ex \@plus .2ex}%
  {\normalfont\Large\bfseries\boldmath}}
\renewcommand\subsection{\@startsection{subsection}{2}{\z@}%
  {-3.25ex \@plus -1ex \@minus -.2ex}{1.5ex \@plus .2ex}%
  {\normalfont\large\bfseries\boldmath}}
\renewcommand\paragraph{\@startsection{paragraph}{4}{\z@}%
  {3.25ex \@plus 1ex \@minus .2ex}{-1em}%
  {\normalfont\normalsize\bfseries\boldmath}}
\makeatother

\newcommand{\Om}{\Omega_{12}}          
\newcommand{\Sig}{\Sigma_{12}}         
\newcommand{\vcross}{\bar v_{\mathrm{cross}}}
\newcommand{\neff}{n_{\mathrm{eff}}}

\DeclareRobustCommand{\datasource}[1]{\textsc{#1.}\enspace}

\newtcolorbox{keybox}[1]{
  enhanced, breakable, colback=white, colframe=white, boxrule=0pt,
  sharp corners, left=0pt, right=0pt, top=3pt, bottom=3pt,
  borderline north={0.7pt}{0pt}{black!70},
  borderline south={0.7pt}{0pt}{black!70},
  title={#1}, fonttitle=\bfseries\small, coltitle=black,
  attach title to upper={\par\smallskip}}

\begin{document}

\title{\textbf{The Noise Is the Signal:\\
Correlated Sampling Error Is Rank-Informative\\
for Proxy Metric Selection}}

\author{Sandro Provenzano\thanks{Email:
\href{mailto:sandro.provenzano@zalando.de}{sandro.provenzano@zalando.de}}\\
\textit{Zalando SE}}

\date{}

\maketitle

\begin{abstract}
North-star metrics such as customer lifetime value are often too slow and
noisy to decide a short A/B test. Teams therefore rely on a proxy metric,
commonly chosen by how closely its effects tracked the north star's across
past experiments. Validating that choice, or any method for making it, is
hard: the only benchmark is the noisy north star, and the number of available
past experiments is limited. In addition, proxy and
north-star effects are estimated on the same customers, so their sampling
errors are correlated. Recent work at major experimentation platforms removes
this shared error as contamination, improving estimates of the true-effect
covariance. Choosing a proxy, however, is a ranking problem, and a better
estimate need not give a better ranking. We measure agreement free of shared
error by estimating the two effects on disjoint random halves of each
experiment's customers. In an archive of 262 experiments and 69 candidate
proxies, the shared error ranks the candidates in a similar order to this
agreement (Spearman correlation $0.65$): it carries information about proxy
quality. The more of it a correction removes, the worse the ranking, because
removal discards part of the signal but leaves the main sources of ranking
noise, the noisy north star and the limited number of experiments, untouched. Archive-calibrated
simulations, in which the correct ranking is known, confirm this even when
every correction receives the true sampling covariance. Held-out real
experiments, evaluated on disjoint customer halves so that shared error cannot
bias the comparison, closely reproduce the predicted ordering (Spearman
correlation $0.93$). Correction can still pay off with more
experiments, but the number needed rises steeply with the north star's noise.
We map this crossover and give platform teams three inexpensive checks for
deciding, from their own archive, whether and how strongly to correct.

\medskip
\noindent\textbf{Keywords:} proxy metrics, surrogate metrics, A/B testing,
meta-analysis, sampling covariance, experimentation platforms, metric
selection, ranking loss
\end{abstract}

\onehalfspacing

\section{Introduction}
\label{sec:intro}

Large experimentation platforms face a common problem. The metrics the
business cares about, such as revenue over a year, retention over six months
or customer lifetime value, move too slowly and too noisily to decide a
short-term test \citep{Tripuraneni2024,Sigerson2026}. Platforms therefore rely on a
\emph{proxy}: a faster metric believed to move in the same direction. Choosing
one well is hard, and the difficulty is circular: the evidence that would
confirm a proxy is the slow, noisy measurement the proxy exists to replace,
the same obstacle that makes surrogate endpoints hard to validate in medicine
\citep{Fleming1996}. The problem remains open, and it has drawn sustained work
from the major platforms over the past decade
\citep{DengShi2016,AnalyticsAtMeta2022,Tripuraneni2024,Bibaut2024,Chou2025,Sigerson2026}.
What is available instead is indirect evidence. One widely used source of it,
and the subject of this paper, is the archive-based workflow: the platform
assembles an archive of past randomized experiments, measures how each
candidate's treatment effects line up with the north star's across that
archive, and ranks candidates by that alignment
\citep{Tripuraneni2024,Zito2025}.

That procedure has a well-known flaw. Both the proxy effect and the north-star
effect in a given experiment are estimated from the same randomized units, so
their estimation errors are correlated. The observed cross-experiment
alignment is therefore the sum of two very different quantities: the
covariance of the true treatment effects across experiments, which is
the quantity of interest, and the covariance of the sampling error
within experiments, which is not. We write these $\Sig$ and $\Om$
respectively. The decomposition itself is not new: \citet{CunninghamKim2022}
write the observed covariance of experiment outcomes
as a true-effect covariance plus a unit-level covariance scaled by sample size,
show that naive surrogacy slopes inherit the second term, and propose netting
it out using A/A moments. \citet{Cunningham2023} derives the bias that $\Om$
induces in regressions of long-run on short-run effects and discusses
adjustments, and practitioner guidance from Meta warns that it would produce a
correlation between proxy and north-star effects even if every experiment were
an A/A test \citep{AnalyticsAtMeta2022}. Estimators that remove it have since
been developed at Netflix \citep{Bibaut2024}, Google
\citep{Tripuraneni2024} and Booking \citep{Gazvoda2024}. These estimators are sound
for their stated purpose, and we use them for exactly that: they calibrate the
true-effect correlations of our simulations and supply three of our five
definitions of true alignment.

We ask a question that, to our knowledge, this literature has not addressed:
does removing $\Om$ improve the ranking? Removing it demonstrably
improves estimation of $\Sig$, the task these estimators are built and
evaluated for. But in choosing a proxy, a platform does not act on the size of
$\Sig$; it acts on its choice of metric. Decision losses are not absent from this literature:
\citet{Chou2025} score decision rules by cumulative reward, and a
parallel literature scores surrogates by the value of the treatment rules they
induce \citep{Yang2023,Xu2026}. What has not been scored is the object the
platform produces: a ranking over a catalog of candidate proxies. That object
deserves scrutiny of its own, because whichever metric tops the ranking goes
on to decide the experiments that follow.

We show that removing $\Om$ can make that ranking worse: when $\Om$ carries
information about which candidates track the north star and the archive is
small relative to the noise in its north star, the subtraction costs more
ranking accuracy than it gains. We also show how a platform can tell from its
own archive whether it is in that regime. On the archive we study, both
conditions hold, and the more of $\Om$ a correction removes, the worse the
ranking gets. The argument has two independent legs: one
concerns what $\Om$ contains, the other what removing it costs.

The first leg is that $\Om$ is not noise for this purpose.
Sampling-error covariance between two metrics measured on the same customers
largely reflects the customer-level relationship between them, a stable
property of the two metrics that also predicts whether the proxy tracks the
north star. We test this with an instrument that does not rely on any of the
estimators being compared. Borrowing a sample-splitting device from the
experimentation literature \citep{CoeyCunningham2019,BibautCrossFold2023}, we
split the customers within each experiment into two disjoint halves, estimate
both the candidate's and the north star's effect on each half, and combine
them across halves so that the sampling errors on the two sides are
independent by construction. The resulting cross-half
agreement requires no variance correction, no estimate of $\Om$ and no
distributional assumption. Across candidates, its Spearman rank correlation
with the within-experiment sampling-error correlation, which is $\Om$ on the
correlation scale, is $0.651$ ($p < 0.0001$). The
association holds on the candidates that were defined before this study
began, and every alternative definition of true alignment we
consider, including three built from the de-noising models themselves,
returns a stronger one (Section~\ref{sec:result}). The quantity the
corrections discard carries much of the rank information the corrections are
trying to recover.

The second leg is that removing $\Om$ is costly in finite samples, for a
reason unrelated to how well $\Om$ is estimated. The corrected statistic is a
difference between a cross-experiment moment, estimated from $K$ experiments
and therefore noisy, and $\Om$, estimated from millions of customers and
therefore very precise. The subtraction leaves the noise in place. When
$\Om$ is largest for the candidates that track the north star best, as it
tends to be on our archive, the subtraction also narrows the gaps between
candidates, and narrows them most at the top of the ranking. We call this
\emph{rank-aligned signal subtraction} (Section~\ref{sec:mechanism}). Its cost
grows with the dose removed. In a simulation built from the archive's own
measured moments, at the archive's size of $K = 262$, pair accuracy (the
probability of ordering two candidates correctly) is $0.854$ for a Deming
correction that adjusts only the proxy-side variance, $0.817$ for the
uncorrected baseline, $0.788$ for a Deming variant that also removes part of
$\Om$, $0.730$ for a total-covariance slope that removes $\Om$ from the
numerator, and $0.643$ for an estimator that removes it and rescales. Every
correction in the simulation receives the exact per-experiment $\Omega_k$ that
generated the data, so what it loses is the cost of removing $\Om$
correctly, not the cost of estimating it poorly (Section~\ref{sec:sim}).
Nor is the ordering an artifact of the simulation: when the same simulated
world is rebuilt with $\Om$ set to zero, the uncorrected baseline falls from
$0.817$ to $0.602$, behind all but one of the corrections. The corrections
work when their premise holds; the first leg tests whether it does.

A third analysis tests the prediction out of sample. We rank candidates on one
set of experiments and score them on a disjoint set; within each scoring
experiment, following \citet{Chou2025}, each decision and its payoff are read
on different customers, so the evaluation is itself free of shared sampling
error. The measures
that keep $\Om$ earn the highest held-out rewards, and the two most aggressive
corrections the lowest. Relative to the uncorrected baseline, six of the seven
measures compared gain or lose held-out reward as the simulation predicts, and
the rank correlation between the simulated and the real gaps to the baseline
is $0.93$ (Section~\ref{sec:horserace}). No single
gap is statistically significant on one archive of this size; the evidence is
the agreement between an out-of-sample ordering and a simulation calibrated
before the comparison was run. The same design shows how much the evaluation
protocol matters: if each decision and its payoff are instead read on the same
customers, shared error inflates the held-out rewards, and the uncorrected
baseline's more than doubles, from $0.249$ to $0.562$.

Two results give the finding relevance beyond this archive. First, the
inversion is a finite-sample effect. Among the estimators of
Figure~\ref{fig:accK}, the most aggressive correction, \texttt{pda}, is
exactly rank-faithful in the limit, yet it ranks worst at every archive size on
our grid, which runs to $K = 1000$. Second, the archive size at which removing
$\Om$ starts to pay rises steeply as north-star reliability falls. At every
reliability rung we simulate at or below $0.60$, the two most aggressive
corrections, \texttt{pda} and \texttt{tc\_rho}, overtake the uncorrected
baseline only once on the grid, by less than half a point at $K = 1000$, and
the three published archives whose size we could verify against a primary
source hold $123$ to $307$ experiments.
Placing them on our map, with our archive's structure held fixed, shows how
far archives of that size sit from the crossover (Section~\ref{sec:map}). The
simulation study of \citet{Bibaut2024} calibrates to a north-star reliability of
$0.976$--$0.999$; ours is $0.104$. Our archive is also smaller than it looks:
measured by leverage, its effective size is about ten experiments for the
median candidate, against a nominal $K = 262$. Much of that concentration
comes from two experiments, but our findings do not rest on them
(Section~\ref{sec:neff}).

The paper makes four contributions. It recasts the premise that justifies
de-noising for ranking, that $\Om$ carries no rank information about $\Sig$,
as an empirical claim that a platform can test with one extra pass over its
customer-level data. It traces the ranking cost of removing $\Om$ to
rank-aligned signal subtraction and separates ranking loss from estimation
loss: the one estimator in our study that is essentially unbiased for the
true-effect correlation trails the uncorrected baseline in pair accuracy by
about ten points (Section~\ref{sec:estvsrank}). It measures, on real
experiments, how much same-sample evaluation flatters a ranking. And it
translates the crossover map into three inexpensive diagnostics (effective
archive size, the cross-half instrument and a contamination-free held-out
reward) that a platform can run on data it already holds before committing to
a correction.

The remainder of this article is organized into eight sections.
Section~\ref{sec:setup} defines the two covariances and
works through the arithmetic for two candidates;
Section~\ref{sec:premise} reviews the literature and describes our archive.
Section~\ref{sec:instrument} develops the first leg on the real archive and
Section~\ref{sec:cost} the second in an archive-calibrated simulation;
Section~\ref{sec:map} places archives on the crossover map and tests the
prediction on held-out real experiments; Section~\ref{sec:practice} sets out
the diagnostics. Section~\ref{sec:limits}
discusses the limits of the evidence, and Section~\ref{sec:conclusion}
concludes. The evidence rests on three main
analyses: the cross-half instrument (Section~\ref{sec:result}), the
counterfactual ablation (Section~\ref{sec:ablation}) and the
contamination-free held-out comparison (Section~\ref{sec:horserace}); the
remaining analyses are controls or robustness checks. Readers mainly interested in practice can go from the
worked example (Section~\ref{sec:example}) to the diagnostics
(Section~\ref{sec:practice}). Because the argument moves between real and
simulated data, the caption of every figure and results table opens by naming
its source: the real archive, the archive-calibrated simulation, a stylized
construction or a combination.

\section{Setup: two covariances}
\label{sec:setup}

Two covariances are easily confused in this literature, and the argument
depends on keeping them apart. This section defines them, fixes the
vocabulary, introduces the estimators we compare and states the loss on which
they are scored.

\begin{table}[t]
\centering
\small
\caption{Terms and distinctions used throughout.}
\label{tab:terms}
\begin{tabular}{@{}P{0.28\columnwidth}P{0.65\columnwidth}@{}}
\toprule
\textbf{Term} & \textbf{Meaning here} \\
\midrule
Sampling error & Estimation error in a treatment effect. Shrinks with the
number of customers in the experiment. \\
\addlinespace
Measurement noise & Error in an individual customer's metric value. Does not
shrink with experiment size. \\
\addlinespace
North-star reliability & The share of the observed cross-experiment variance
in $\hat\tau^{y}$ that is true-effect rather than sampling-error variance,
$\mathrm{var}(\tau^{y})/\mathrm{var}(\hat\tau^{y})$, the reliability ratio of
the measurement-error literature \citep{Fuller1987}. It runs from zero, where
the measured effects are pure noise, to one, where they are measured without
error. \\
\addlinespace
Archive size $K$ & The number of experiments in the archive. \\
\addlinespace
Effective archive size $\neff$ & The number of experiments' worth of
independent leverage the archive supplies once unequal precision is accounted
for: the inverse-Herfindahl count $(\sum_k d_k^{2})^{2}/\sum_k d_k^{4}$ of the
proxy-effect deviations $d_k$, which falls far below $K$ when a few
experiments dominate the fit (Section~\ref{sec:neff}). \\
\addlinespace
Pair accuracy & The probability that two candidates are ordered correctly. A
ranking loss, not an estimation loss. \\
\addlinespace
``Correlation'' & Always qualified: either the
\emph{within-experiment sampling-error correlation} or the
\emph{cross-experiment true-effect correlation}. \\
\bottomrule
\end{tabular}
\end{table}

\subsection{The estimand and the nuisance}

Index experiments by $k = 1,\dots,K$. Each experiment yields an estimated
treatment effect on a proxy candidate, $\hat\tau^{x}_k$, and on the north star,
$\hat\tau^{y}_k$, both computed from the same randomized customers. Write
\begin{equation}
\begin{pmatrix}\hat\tau^{x}_k\\[2pt] \hat\tau^{y}_k\end{pmatrix}
=
\begin{pmatrix}\tau^{x}_k\\[2pt] \tau^{y}_k\end{pmatrix}
+
\begin{pmatrix}e^{x}_k\\[2pt] e^{y}_k\end{pmatrix},
\qquad
\mathrm{Var}\!\begin{pmatrix}e^{x}_k\\ e^{y}_k\end{pmatrix}
= \Omega_k =
\begin{pmatrix}\omega^{xx}_k & \omega^{xy}_k\\ \omega^{xy}_k & \omega^{yy}_k\end{pmatrix}.
\end{equation}
The true effects $(\tau^{x}_k,\tau^{y}_k)$ vary across experiments with
covariance $\Sig = \mathrm{Cov}(\tau^{x},\tau^{y})$. This is the
\emph{cross-experiment true-effect covariance}: the estimand. Divided by the
standard deviations $\sigma_x$ and $\sigma_y$ of the true effects, it becomes
the \emph{cross-experiment true-effect correlation}
$\rho = \Sig/(\sigma_x\sigma_y)$, the truth against which our simulations
score every ranking. The
within-experiment sampling errors have covariance $\omega^{xy}_k$; we write
$\Om$ for the archive-average of $\omega^{xy}_k$, and call it the
\emph{within-experiment sampling-error covariance}. The observed
cross-experiment covariance of the estimates is
\begin{equation}
\mathrm{Cov}(\hat\tau^{x},\hat\tau^{y}) \;=\; \Sig \;+\; \Om .
\label{eq:decomp}
\end{equation}
The decomposition follows from the two-level model of bivariate
random-effects meta-analysis, which keeps the within-study correlation between
two estimated effects explicit \citep{vanHouwelingen2002,Riley2009} and
underlies the meta-analytic evaluation of surrogate endpoints
\citep{DanielsHughes1997}; \citet{CunninghamKim2022} derive it for online
experiments. Equation~\eqref{eq:decomp} also states the problem. If the target is $\Sig$, then
$\Om$ is a bias term, and the literature's response, to estimate $\Om$ and
subtract it, is the right one. Our question is what happens to a
ranking over candidates $x$ when that subtraction is performed at
realistic $K$. Table~\ref{tab:terms} defines the terms we use throughout.

\subsection{A worked example: why an exact correction can rank worse}
\label{sec:example}

\begin{keybox}{Example: two candidates}
\small
Two candidates, $A$ and $B$. $A$'s true effects covary with the north star's at
$\Sig^{A} = 1.0$; $B$'s at $\Sig^{B} = 0.8$. The correct ranking is
$A \succ B$. Because $A$ is more tightly related to the north star at the
customer level, it also has the larger sampling-error covariance:
$\Om^{A} = 1.0$, $\Om^{B} = 0.4$.

\textbf{Uncorrected.} Observed covariances are $2.0$ and $1.2$. Ranking:
$A \succ B$ (correct).

\textbf{Corrected, with $\Om$ known exactly.} $2.0 - 1.0 = 1.0$ and
$1.2 - 0.4 = 0.8$. Ranking: $A \succ B$ (correct and unbiased).

\textbf{Corrected, at finite $K$.} The observed covariances are not known:
each is estimated from $K$ experiments with some standard error $s$. Because
$\Om$ is estimated from millions of customers rather than from $K$ experiments,
subtracting it leaves $s$ essentially unchanged. What it does shrink is the gap
that $s$ must resolve, from $0.8$ to $0.2$. To come out right as often as the
uncorrected ranking, the corrected one needs a standard error four times
smaller, and hence sixteen times as many experiments; at modest $K$ it is
reversed a substantial share of the time. This happens even though the
subtraction is exact.
\medskip

\noindent The correction trades a bias that is rank-preserving here for
a variance that is not. Whether that trade is worth it depends on whether
$\Om$ and $\Sig$ are positively rank-associated, and on $K$. Both can be
measured, and the rest of the paper measures them.
\end{keybox}

\noindent The example does not imply that the correction is invalid: with
$\Om$ known, it is exactly right. Nor does it treat $\Om$ and $\Sig$ as the
same object, or depend on $\widehat{\Om}$ being poorly estimated, since the
subtraction in the example is exact. What it shows is that when $\Om$ and
$\Sig$ are positively rank-associated, an unbiased correction can rank worse
than a biased estimator at finite $K$. This bias--variance trade-off is
invisible in the estimation framing, because under an estimation loss the
correction is better.

\subsection{What \texorpdfstring{$\Om$}{\textOmega12} contains}
\label{sec:whatomega}

The example turned on one assumption: that the candidate more tightly related
to the north star at the customer level also has the larger sampling-error
covariance. The argument of this paper turns on that assumption, so we begin
with what $\Om$ is.

The literature's framing of $\Om$ as contamination is natural if one thinks of
it as an artifact of finite samples. For two metrics measured on the same
customers under simple randomization, $\omega^{xy}_k$ is the customer-level
covariance of the two metrics in each arm divided by that arm's number of
customers, summed over the two arms. Its sign and relative magnitude across candidates
are therefore governed by the customer-level association between the proxy and
the north star: a structural property of the metric pair, not an artifact of
any one experiment's sample size.

This is why $\Om$ is not obviously discardable when the task is ranking. A
candidate strongly related to the north star at the customer level will tend to
have a large $\omega^{xy}$. Whether its true effects also tend to covary
with the north star's is a separate question, and it is the one the example
assumed an answer to. The two channels are distinct, and conflating them is
the error the de-noising literature corrects; the meta-analytic validation of
surrogate endpoints likewise keeps them apart, as individual-level and
trial-level association \citep{Buyse2000}. Neither channel supports a
causal claim about the proxy: a customer-level association between two metrics
does not license the inference that intervening on one moves the other, which is the inferential gap the surrogacy literature addresses
\citep{Fleming1996,VanderWeele2013}.

Nor is the direction of the relationship obvious a priori. The
literature's working prior runs the other way. \citet{CunninghamKim2022}
observe that desirable outcomes usually covary positively across users
while treatments frequently trade them off against one another, and
\citet{Bibaut2024} motivate their correction with the same picture: clicks and
conversions correlated across users, click-bait treatments that raise one and
lower the other. On that prior, $\Om$ and $\Sig$ are anti-aligned across
candidates, so the first leg of our argument fails and removing $\Om$ is the
right choice for ranking as well as for estimation. Whether the prior holds is
an empirical question about a particular archive and catalog.
Section~\ref{sec:instrument} measures how much rank information the first
channel carries about the second rather than assuming an answer.

\subsection{The estimators and a note on naming}
\label{sec:estimators}

Table~\ref{tab:dose} lists the eight estimators behind the main results,
together with one reference statistic. They differ chiefly in how much
of $\Om$ they remove, and the results below are organized by that dose. Three
labels recur. \emph{Uncorrected baselines} leave $\Om$ in place. \emph{Mild
corrections} remove none or only part of it while correcting a variance.
\emph{Full de-noising} removes it entirely. The reference statistic,
\texttt{obs\_corr}, is not on this scale: it is $\Om$, expressed as a
correlation.

\begin{table}[t]
\centering
\small
\caption{The eight estimators behind the main results and the reference
statistic \texttt{obs\_corr}, grouped by how much of the within-experiment
sampling-error covariance $\Om$ they remove; results throughout are organized
by this dose. \texttt{pda} and \texttt{tc\_rho} remove the same amount and
differ only in what they divide by afterwards; $\hat\sigma_x$ and
$\hat\sigma_y$ are de-noised estimates of the standard deviations of the
candidate's and the north star's true effects across experiments. Formulas,
implementations and sources are in Appendix~\ref{app:estimators}.}
\label{tab:dose}
\begin{tabular}{@{}lll@{}}
\toprule
\textbf{Estimator} & \textbf{Treatment of $\Om$} & \textbf{Scale} \\
\midrule
\multicolumn{3}{@{}l}{\emph{Uncorrected baselines}} \\
\texttt{ols\_r}            & kept                                   & correlation \\
\texttt{ols\_slope}        & kept                                   & slope \\
\addlinespace
\multicolumn{3}{@{}l}{\emph{Mild corrections}} \\
\texttt{deming\_slope}     & kept; $x$-variance corrected           & slope \\
\texttt{deming\_cov\_slope}& partially removed                      & slope \\
\addlinespace
\multicolumn{3}{@{}l}{\emph{Full de-noising}} \\
\texttt{tc\_slope}         & removed from numerator                 & slope \\
\texttt{pda}               & removed; $\div\,\hat\sigma_x$          & score \\
\texttt{tc\_rho}           & same dose; also $\div\,\hat\sigma_y$   & correlation \\
\texttt{re\_rho}           & modeled away                          & correlation \\
\addlinespace
\multicolumn{3}{@{}l}{\emph{Reference statistic}} \\
\texttt{obs\_corr}         & \emph{is} $\Om$                        & correlation \\
\bottomrule
\end{tabular}
\end{table}

Seven of the eight estimators are established methods. \texttt{ols\_r} and
\texttt{ols\_slope} are the correlation and slope of an uncorrected
least-squares fit of $\hat\tau^{y}$ on $\hat\tau^{x}$. \texttt{deming\_slope}
is a Deming regression \citep{Deming1943}, an errors-in-variables fit with
known, experiment-specific error variances, which corrects for the proxy's
sampling variance; \texttt{deming\_cov\_slope} is York's extension to
correlated errors \citep{York1968}, which also takes the sampling covariance
into account. \texttt{tc\_slope} and \texttt{tc\_rho} rest on the
total-covariance estimator of \citet{Bibaut2024}, which subtracts the
archive-average sampling covariance and variances from the cross-experiment
covariance and variances of the estimates. This gives a de-noised covariance
$\widehat{\Sig}$ and de-noised standard deviations $\hat\sigma_x$ and
$\hat\sigma_y$. \texttt{tc\_slope} is the slope \citeauthor{Bibaut2024} derive
from these moments, $\widehat{\Sig}/\hat\sigma_x^{2}$, and \texttt{tc\_rho} the
corresponding correlation, $\widehat{\Sig}/(\hat\sigma_x\hat\sigma_y)$.
\texttt{re\_rho} is the true-effect correlation of a bivariate random-effects
meta-analysis \citep{vanHouwelingen2002}, fitted by restricted maximum
likelihood; its square is the trial-level $R^{2}$ that medicine uses to
validate surrogate endpoints, in the version computed from treatment effects
alone \citep{Buyse2000}, and \citet{Tripuraneni2024} and \citet{Gazvoda2024}
develop Bayesian hierarchical versions of the model for proxy metrics.

The eighth estimator, \texttt{pda}, is our own construction, the
\emph{partially de-noised alignment} $\widehat{\Sig}/\hat\sigma_x$. It
removes exactly as much of $\Om$ as \texttt{tc\_rho} but omits the division
by $\hat\sigma_y$. Because $\hat\sigma_y$ is common to all candidates, that
division cannot reorder them, and wherever \texttt{tc\_rho} is defined the
two rank the candidates almost identically. What the division can do is leave
\texttt{tc\_rho} without a value: the north-star variance is poorly
determined, and its de-noised estimate can be negative, which leaves
\texttt{tc\_rho} undefined, or so small that \texttt{tc\_rho} falls outside
$[-1, 1]$ (Section~\ref{sec:tcrho}). \texttt{pda} is immune to both
failures, so comparing it with the uncorrected baseline isolates the cost of
removing $\Om$, and comparing it with \texttt{tc\_rho} the cost of
estimating that variance (Section~\ref{sec:dose}).

Appendix~\ref{app:estimators} gives the formulas and implementation details.
We implemented all estimators in Python, so that the whole pipeline runs in a
single language. Where established reference software exists, our
implementations agree with it: on the real archive, \texttt{re\_rho} matches
the R package \texttt{metafor} \citep{Viechtbauer2010}, and the two Deming
slopes match \texttt{deming} \citep{Therneau2024} and \texttt{IsoplotR}
\citep{Vermeesch2018}, each to within $3\times10^{-5}$ and with identical
candidate rankings; and the
total-covariance estimator, run on the simulation design of
\citet{Bibaut2024}, matches the authors' published code to within
$1.7\times10^{-12}$ on the estimated slope.

\paragraph{Baseline and observational measures.} \texttt{ols\_r} is the
\emph{uncorrected baseline}: the analysis a platform runs when it applies no
correction. \texttt{obs\_corr}, as we call it in the simulations, is the
within-experiment sampling-error correlation,
$\Om/(\bar\omega^{xx}\bar\omega^{yy})^{1/2}$, where $\bar\omega^{xx}$ and
$\bar\omega^{yy}$ are the archive-average sampling variances of the candidate
and the north star; it uses no effect estimates at all. On the real archive
the same statistic is the horizontal axis of the instrument
(Figure~\ref{fig:rankrank}), and in the held-out comparison
(Section~\ref{sec:horserace}), computed on the ranking half, it appears as
\texttt{obs\_within\_experiment\_corr}, next to its unstandardized numerator,
\texttt{obs\_within\_experiment\_cov}. ``Observational'' refers to the
moments a statistic uses (within-experiment, customer-level
association), not to the design: every experiment in the archive is
randomized, and the term is unrelated to the observational surrogate index of
the surrogate literature, which is fitted on non-experimental data
\citep{Athey2025,Sigerson2026}.

\subsection{The loss function}
\label{sec:loss}

The literature on surrogate and proxy metrics has long distinguished the
statistical validity of a surrogate from its usefulness for decisions
\citep{Fleming1996,Frangakis2002,VanderWeele2013}. The distinction remains
central to current practice. At a workshop with 26 experts from fifteen leading
online platforms and four universities, the tension between unbiased estimation and
decision-making was a recurring theme, and participants noted that getting
launch decisions right does not imply estimating the long-run effect without
bias, because the former ``usually implies a different loss function'' than
the latter \citep{Sigerson2026}. The same separation has recently been
formalized for individualized treatment rules \citep{Xu2026}, and for rankings
it is classical: in hierarchical models, the estimates that are optimal for
each unit's parameter need not be optimal for ordering the units
\citep{ShenLouis1998,Lin2006}. We therefore
evaluate estimators under the losses a platform incurs when it shortlists
candidates.

Our primary loss is \emph{pair accuracy}: over random pairs of candidates,
the probability the estimator orders them as the truth does. It is scale-free
and robust to monotone transformations, and it directly reflects the
shortlisting task. Absent ties and unscored candidates, it is Kendall's rank correlation
between scores and truth \citep{Kendall1938}, rescaled from $[-1,1]$ to
$[0,1]$. Two conventions apply. An exact tie on either side counts as wrong; on
continuous scores ties have probability zero, so this matters only in the
ablation (Section~\ref{sec:ablation}). A pair the estimator cannot order,
because it is undefined for one of the candidates, also counts as wrong rather
than being dropped; since this matters for only one estimator, we also report
accuracy over defined pairs (Sections~\ref{sec:threelosses}
and~\ref{sec:tcrho}). We report two secondary
losses. \emph{Top-5 regret} is the mean true alignment of the true top five
minus the mean true alignment of the five the estimator selects; when an
estimator cannot rank a slot, the slot is filled from the candidates it did
not score, so failing to return a score is not rewarded. It is measured in
units of the true correlation, and it weights the head of the ranking, where
shortlisting decisions tend to be made. \emph{Top-5 recovery} is the share
of the true top five that appears in the estimated top five. We also report
estimation losses (bias and RMSE for $\rho$), not as a check on the ranking
results but as a deliberate contrast: they order the estimators differently,
and that divergence is itself a result (Section~\ref{sec:estvsrank}).

The choice between a ranking loss and an estimation loss is central to this
paper: an estimator chosen for its small bias need not rank best, which is how
a literature that optimizes an estimation criterion can produce estimators
that rank worse (Section~\ref{sec:estvsrank}).

\section{The premise and where it has been validated}
\label{sec:premise}

This section sets out what the de-noising literature establishes and where it
has been validated, and then describes our archive.

\subsection{What the literature establishes}

That $\Om$ biases naive estimators of $\Sig$ is well established: the result
has been derived independently several times and demonstrated on platform
data. Table~\ref{tab:t1} summarizes what each source establishes and the
setting in which it was validated.

\begin{table}[t]
\centering
\small
\caption{The de-noising literature: what each source establishes about
correlated sampling error, and the setting in which it is validated.}
\label{tab:t1}
\begin{tabular}{@{}P{0.20\textwidth}P{0.45\textwidth}P{0.29\textwidth}@{}}
\toprule
\textbf{Source} & \textbf{What it establishes about correlated sampling error}
& \textbf{What it validates it on} \\
\midrule
\citet{CunninghamKim2022} &
Observed cross-experiment covariance is true-effect covariance plus unit-level
covariance over sample size; naive surrogacy slopes inherit the second term,
which can be netted out with A/A moments &
Gaussian analysis; illustrative A/A scatter \\[2pt]
\citet{Bibaut2024} (Netflix) &
Measurement noise biases naive covariance estimators; a total-covariance
estimator recovers the treatment-effect covariance under a common-$\Om$
assumption, and LIMLK recovers the surrogate-to-outcome structural parameter,
efficiently but inconsistently under direct effects &
Bias in simulation; a platform application \\[2pt]
\citet{Cunningham2023} &
Naive short-/long-run regressions are biased because long-run noise on the
left-hand side is correlated with short-run measurement noise &
Analytic decomposition; practitioner essay \\[2pt]
\citet{Gazvoda2024} (Booking) &
Observed correlation is contaminated by correlated sampling error rather than
reflecting the correlation of true treatment effects &
500 simulated corpora of \textbf{20} experiments; DGP matches their model \\[2pt]
\citet{Chou2025} (Netflix) &
The naive estimator depends on the covariance of measurement error $\Om$ and
can rank a noisy bad proxy above a good proxy &
Analytic Gaussian example; \textbf{123} A/B tests, rule reward \\[2pt]
\citet{AnalyticsAtMeta2022} &
Even if every experiment were an A/A test, correlated sampling error would
produce a proxy--ground-truth correlation equal to the unit-level one;
recommends multivariate shrinkage to help account for it &
Practitioner playbook \\[2pt]
\citet{Tripuraneni2024} (Google) &
A hierarchical model denoises observed treatment effects and disentangles
latent cross-experiment variation from within-experiment noise &
\textbf{307} A/B tests, proxy score \\
\bottomrule
\end{tabular}
\end{table}

Two features of the right-hand column matter. First, validation is largely on
estimation targets (recovery of $\Sig$ or of a downstream rule reward)
rather than on the ranking loss of Section~\ref{sec:loss}. This is the gap we
address, and it does not depend on reliability: to our knowledge no paper in
this literature has evaluated a de-noising correction by the finite-sample
loss it incurs when ordering a fixed catalog of candidate proxies. The
nearest approaches score a single decision rule's return \citep{Chou2025}, a
proxy score against a held-out criterion \citep{Tripuraneni2024} or the
sensitivity--directionality frontier among candidates while naming the
contamination and leaving it uncorrected \citep{Zito2025}. Second, where
simulation is used, the reliability of the simulated north star is rarely
reported, so it is hard to know from the outside which regime a given result
was established in. We reconstructed the DGP of \citet{Bibaut2024} from the
authors' published code and found its north-star reliability to lie between
$0.976$ and $0.999$, a regime that no metric in our archive occupies, least of
all our north star (Figure~\ref{fig:ladder}).

\paragraph{The wider proxy-metric literature.} De-noising sits inside a larger
effort to make short-horizon measurement stand in for long-horizon value. One
strand builds proxies chiefly for power: learning metric combinations
that maximize statistical power with respect to the north star
\citep{Jeunen2024}, or exploiting in-experiment data to reduce variance in
sparse and delayed outcomes \citep{Deng2023}. Both screen for fidelity as well
(Jeunen and Ustimenko penalize directional disagreement with the north star,
and Deng et al.\ validate alignment on the 32 of 141 experiments in which the
delayed outcome moved significantly), but the fidelity check is run on
observed effects, so it inherits $\Om$ in the same way as the
correlation does.
A second strand estimates long-run effects from short-run signals: by
combining short-horizon outcomes into a surrogate index under a Prentice-style
conditional-independence assumption \citep{Prentice1989,Athey2025}, whose decision value has
been audited on 200 A/B tests at Netflix \citep{Zhang2024}; by meta-mediation
across 190 Etsy experiments under explicitly stated assumptions to recover a
dose-response function \citep{Wang2020}; or by modeling the response path of
long-term treatments over time \citep{Huang2026}. Practitioner guidelines for
qualifying surrogate metrics in this mode are also documented
\citep{Duan2021,AnalyticsAtMeta2022}. A third and older strand in biostatistics
asks when a surrogate is licensed to replace an endpoint at all
\citep{Prentice1989,Fleming1996,Frangakis2002,VanderWeele2013}, and, closest in spirit to
the many-experiments setting here, how to validate one by pooling across
trials \citep{DanielsHughes1997,Buyse2000}, including how to measure the risk
of a surrogate paradox \citep{Elliott2015}. Our question is more specific
than any of these and largely complementary to them: given a fixed catalog of
candidates and a fixed archive, does removing $\Om$ produce a better
ordering? The estimation results of Table~\ref{tab:t1} do not answer
it: a correction that recovers each candidate's alignment more accurately need
not order the candidates more accurately (Section~\ref{sec:example}).

\subsection{The archive and its north star}
\label{sec:archive}

Our archive comprises 262 randomized experiments run on a single
consumer platform, spanning two customer-experience domains (184 experiments
in one, 78 in the other), with a median of 2.25 million customers assigned
per arm. We evaluate 69 proxy candidates: 14
individual metrics and 55 composite indices. Forty-nine of them, the
individual metrics and 35 of the indices, were defined before any of the
analysis in this paper, and we refer to them as pre-declared. As our north-star
metric, we use short-term profit measured over the experiment window plus the
four weeks following it.

\paragraph{How the other 20 candidates were chosen.} The
pre-declared indices all combine browsing- and ordering-funnel metrics,
leaving a whole region of the design space unexamined, so we enumerated all
2{,}497 equal-weight combinations of between two and six individual metrics
and added 20, picked by bootstrap front-stability on a Pareto front of
\emph{sensitivity} (share of experiments in which the effect is significant)
against \emph{directionality}, the two metric properties this literature has
long traded off against one another \citep{DengShi2016,Zito2025}. The
directionality axis raises a selection concern. It is the uncorrected
observed cross-experiment correlation, the same quantity that serves as this
paper's baseline measure, and the selection was run on the full 278-experiment
extraction, which contains the 262 analyzed here. Those 20 candidates were
therefore chosen using data that later results also use.

The resulting bias could run either way. The observed correlation is the sum of
a latent-alignment term and a sampling-covariance term
(Equation~\eqref{eq:decomp}), so selecting on it makes the two
negatively dependent within the selected set (Berkson's paradox;
\citealp{Berkson1946}), which weakens the association we test among the selected
candidates. Selection also places the selected candidates high on both, which
strengthens the association once they are pooled with the rest. We therefore
control for the selection directly: every headline result is also reported on
the 49 pre-declared candidates alone (Section~\ref{sec:result},
Table~\ref{tab:controls}), and the conclusion is unchanged.
Appendix~\ref{app:selection} gives the full selection rule and compares the
two groups.

\paragraph{Randomization integrity and the analyzed population.} We ran
sample-ratio-mismatch tests \citep{Fabijan2019} on every experiment and
excluded those that failed at $p < 0.001$; among the remaining 262, the tests
flag no more experiments than chance would. The analysis is complete-case,
which changes the estimand: effects are average effects on customers with
complete metric records, not on all assigned customers. It does not break
randomization, because attrition is symmetric across arms: in no experiment
does the share of customers with complete records differ between the arms by
more than $0.38$ percentage points. The extraction's outcome filter, which
drops any (experiment, base metric) cell whose absolute measured effect
exceeds $0.10$ control standard deviations, applies to $22$ of $18{,}078$
candidate--experiment cells, or $0.12\%$, and it is concentrated rather than
diffuse: all $22$ come from a single experiment, and all trace to a single
base metric plus the $21$ composites that contain it. Twenty-two candidates
therefore lose one of their $262$ experiments and the other $47$ lose none.

\begin{figure}[t]
\centering
\includegraphics[width=0.63\textwidth]{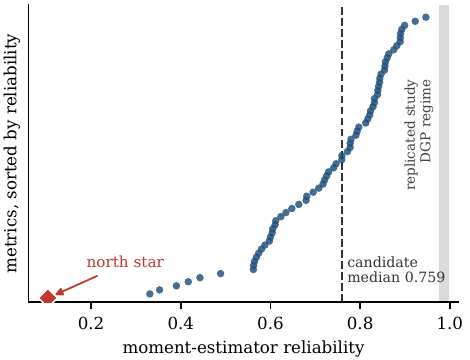}
\caption{\datasource{Real archive} The north star is the least reliably
measured metric in the archive. Each point is one of the 70 metrics (69
candidates and the north star), sorted by reliability. Dashed line:
candidate median. Shaded band: north-star reliability in the simulation design
of \citet{Bibaut2024}, reconstructed from the authors' published code.}
\label{fig:ladder}
\end{figure}

Figure~\ref{fig:ladder} shows the property of the archive that matters most
for what follows. The north star's reliability (the share of its observed
cross-experiment effect variance that is true-effect variance) is $0.104$, the
lowest of all 70 metrics and a factor of $7.3$ below the candidate median of
$0.759$. A
random-effects estimate puts it at $0.125$, and the bootstrap interval around
the moment estimate, $[-0.269,\,0.322]$, comfortably includes zero. We see this less as a
peculiarity of our platform than as a consequence of what north-star status
means: metrics are promoted to it because they measure something slow and
consequential, and slow, consequential outcomes are measured poorly over the
short horizon an experiment allows. Experts from fifteen online platforms identify this as the
defining difficulty of long-term evaluation \citep{Sigerson2026}. We cannot
show that every platform's north star is as unreliable as ours, and we do not
claim it. What we can show is that this north star lies far outside the
regime in which the total-covariance correction was validated \citep{Bibaut2024}. That regime, a
north-star reliability of $0.976$--$0.999$ in the published simulation design,
is the shaded band in Figure~\ref{fig:ladder}, and no metric in our archive
reaches it.

\section{Is \texorpdfstring{$\Om$}{\textOmega12} rank-informative? A cross-half test}
\label{sec:instrument}

This is the first of the paper's three main analyses. It asks whether $\Om$
carries rank information about true alignment on this archive, and answers
without fitting a de-noising model, estimating an effect model or subtracting
anything.

\subsection{The circularity problem}

To ask whether $\Om$ carries rank information about true alignment, one needs
a definition of true alignment. On real data every such definition is itself
an estimate, and the obvious candidates are the estimators being compared. If
we define truth by a random-effects fit, we have assumed the de-noising model;
if we define it by the uncorrected baseline, we have assumed its negation.
Either way the test is circular, because the choice of truth already contains
the answer.

This is the central methodological obstacle. We address it in three steps: we
(i) state the properties a non-circular truth definition must have, (ii)
construct one and (iii) report the answer under all the definitions,
circular ones included, so that the reader can see how much the conclusion
depends on the choice.

A truth definition is admissible for this test if it uses none of the inputs
that distinguish the estimators being compared: no estimate of
within-experiment sampling variance, no estimate of $\Om$ and no parametric
model of the cross-experiment effect distribution.

\subsection{The cross-half agreement instrument}

Within each experiment, we split the customers at random into two disjoint
halves and estimate the candidate's effect and the north star's effect
separately on each half. Because the two halves contain different
customers, the sampling errors in the two estimates are independent by
construction. The sample covariance of the two cross-half series across
experiments therefore estimates the covariance of the true effects with
nothing to subtract: $\mathrm{Cov}(\hat\tau^x_1, \hat\tau^y_2) =
\mathrm{Cov}(\tau^x,\tau^y) + \mathrm{Cov}(e^x_1, e^y_2) =
\mathrm{Cov}(\tau^x,\tau^y)$, because the error cross-term is zero by design
rather than by assumption. We center each series at its own cross-experiment
mean, so that the statistic is a covariance, not a raw second moment,
into which a nonzero average effect would otherwise enter. The same
construction, applied to a candidate against itself across halves and to the
north star against itself, supplies the two variances the ratio needs. We use
both cross directions and average them, so each experiment enters twice and
the statistic does not depend on which half is labeled first. The ratio
applies Spearman's correction for attenuation \citep{Spearman1904}, with the
two halves of each experiment serving as independent repeated measurements
of the same effects. We refer to the resulting per-candidate statistic as
\emph{cross-half agreement}.

Three properties make it admissible. It subtracts no variance estimate, so it
cannot inherit the variance corrections' assumptions. It uses no estimate of
$\Om$ (fold independence replaces the subtraction), so it cannot mechanically
reproduce $\Om$. And it makes no distributional assumption about $\tau$, so it
does not embed the random-effects model.

\paragraph{Lineage and terminology.} Splitting one experiment into
independent sub-experiments so that an estimate and the quantity it is compared
against no longer share sampling noise is not new. \citet{CoeyCunningham2019} split 226 Facebook A/B tests into disjoint
sub-experiments so that one side could serve as an unbiased regression target
for the other, and cross-fold moment constructions serve the same purpose in
the many-weak-experiments literature \citep{BibautCrossFold2023,Chou2025}. We borrow both the split and Spearman's
correction, and we use them to evaluate estimators rather than to build a
new one. We also use \emph{instrument} throughout in
the ordinary sense of a measuring device, chosen here for what it does not
depend on, and not in the instrumental-variables sense in
which past experiments instrument for a surrogate
\citep{PeysakhovichEckles2018,BibautCrossFold2023}: there is no exclusion
restriction here and nothing is estimated by two-stage least squares.

\paragraph{Sampling noise in the instrument.} Each side of the split carries
half the customers, so cross-half agreement is considerably noisier than the
model-based estimators, and it is a ratio of unbiased-but-noisy moment
estimates rather than a correlation, so it is not bounded by one. Its values
here span $0.153$ to $1.431$, with 16 of 69 candidates exceeding $1$. We
therefore use the statistic only ordinally and do not interpret its values as
correlations. Every test below is a rank test, and
Figure~\ref{fig:rankrank} is plotted on ranks for this reason. Every
candidate is measured in every experiment, so each statistic rests on
essentially the whole archive: the outcome filter of Section~\ref{sec:archive},
applied within each half, removes at most two of the 262 experiments from any
candidate's series. The ratio is defined only when both cross-half variances
are positive, which holds for all 69 candidates.

\subsection{Result}
\label{sec:result}

\begin{figure}[t]
\centering
\includegraphics[width=0.63\textwidth]{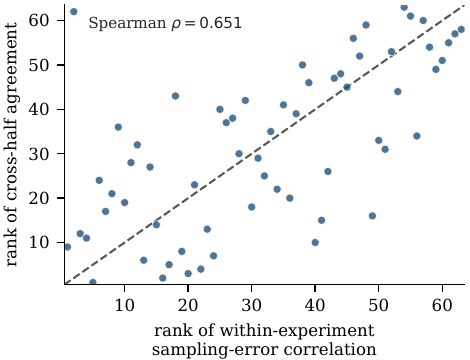}
\caption{\datasource{Real archive} The within-experiment sampling-error
correlation and cross-half agreement largely agree on the ranking of the
candidates. Each point is one of the 63 candidates scored under every truth
definition of Table~\ref{tab:truth}, placed by its rank on the two statistics.
Ranks are plotted because cross-half agreement can exceed one. Dashed line:
perfect rank agreement. Spearman
$\rho = 0.651$, $p < 0.0001$, 95\% bootstrap CI $[0.43,\,0.80]$;
$\rho = 0.686$ on all 69 candidates.}
\label{fig:rankrank}
\end{figure}

Figure~\ref{fig:rankrank} gives the answer. The within-experiment
sampling-error correlation and the model-free instrument rank the
candidates in substantial agreement: Spearman $\rho = 0.651$ on the 63
candidates that every truth definition in Table~\ref{tab:truth} scores,
$p < 0.0001$, with a 95\% bootstrap interval of
$[0.43,\,0.80]$; every one of 10{,}000 bootstrap resamples was positive. On all
69 candidates the figure is $0.686$ with interval $[0.52,\,0.80]$, so the
restriction does not drive the result. Nor does the result depend on how the
instrument is scoped: Appendix~\ref{app:designs} re-computes it on the earlier
and later halves of the archive ($0.709$ and $0.657$) and on each product
domain separately ($0.597$ and $0.907$), and finds it substantial
throughout.

It does not depend on candidate selection either. Restricted to the 49
pre-declared candidates (dropping every candidate chosen by the
front-stability rule of Section~\ref{sec:archive}, and with it every candidate
chosen using an uncorrected observed correlation), the association is
$\rho = 0.626$ ($p < 0.0001$; $0.599$ on the 47 of those
that every truth definition scores), with a 95\% bootstrap interval of
$[0.36,\,0.81]$ and all but one of 10{,}000 resamples positive. Among the 20
selected candidates on their own it remains clearly positive, $\rho = 0.571$
($p = 0.008$). This shows that selection did not produce the headline
association, because removing the selected candidates leaves it strongly
positive. Selection is not entirely without effect, however: in the held-out
comparison of Section~\ref{sec:horserace}, the uncorrected baseline scores
lower once the selected candidates are removed.

\paragraph{Dependence among candidates.}
Fifty-five of the candidates are equal-weight composites of the same 14 base
metrics, so the candidate axis does not contain 69 independent draws, and
every interval and $p$-value in this section is correspondingly optimistic.
The most stringent check runs the instrument on the 14 individual
base metrics alone. The
association stays positive at about three-quarters of the pooled size,
$\rho = 0.499$, but at $n = 14$ it is no longer separated from zero: 95\%
interval $[-0.10,\,0.90]$, $p = 0.069$, $95.4\%$ of resamples positive. We
read this as the resolution limit of the instrument rather than as a contrary
result: the point estimate falls only moderately and stays positive, and what
is lost is precision, as expected when $n$ falls from 69 to 14. It is also why
the instrument addresses only one leg of the argument: it shows that $\Om$ is
rank-informative on this archive, while the held-out comparison of
Section~\ref{sec:horserace}, whose unit of resampling is the experiment
rather than the candidate, measures what that rank information is worth in a
decision.

\paragraph{The same question on the covariance scale.} Both sides of the
instrument can be formed without dividing by anything: the within-experiment
sampling covariance $\widehat{\Om}$ itself against the raw cross-half
product moment. That version subtracts nothing and standardizes nothing, so it
cannot inherit a shared-denominator artifact from the two variances; it is
also far noisier, because it leaves candidates on incommensurable scales. It
returns $\rho = 0.336$ on all 69 candidates ($p = 0.005$, $[0.08,\,0.55]$) and
$\rho = 0.390$ on the 63 ($p = 0.002$, $[0.15,\,0.59]$). The sign and the
separation from zero hold, but the magnitude does not, and among the 20
selected candidates alone the covariance-scale association is flat
(Appendix~\ref{app:selection}). We
treat the correlation-scale figure as the ranking-relevant one, because
candidates are ranked on standardized scores and not on raw moments, but we
report both and use both in Section~\ref{sec:mechanism}, where the gap between
them prevents the mechanism analysis from settling the quantitative question.
For scale, the spread of $\widehat{\Om}$ across candidates is $1.17$
times that of the cross-half moment it is compared with, so the two quantities
vary on comparable scales.

Table~\ref{tab:truth} reports the same association under the other four truth
definitions, including the three built from de-noising models, which are
circular.

\begin{table}[t]
\centering
\small
\caption{\datasource{Real archive} Rank association between the
within-experiment sampling-error
correlation and ``true'' alignment under five definitions of truth
($n = 63$ candidates defined under all five). Rows marked \checkmark\ build
truth from a de-noising model and are therefore circular in the direction
that would favor the corrections.}
\label{tab:truth}
\begin{tabular}{@{}lcc@{}}
\toprule
\textbf{Definition of true alignment} & \textbf{De-noised?} & \textbf{Spearman} \\
\midrule
Cross-half agreement (ours)            & ---          & $0.651$ \\
Random effects, covariance set to zero & \checkmark   & $0.760$ \\
Random effects (full)                  & \checkmark   & $0.697$ \\
Total-covariance estimator             & \checkmark   & $0.666$ \\
Uncorrected baseline \texttt{ols\_r}   & ---          & $0.886$ \\
\bottomrule
\end{tabular}
\end{table}

Every definition returns a strong positive association, from $0.651$ to
$0.886$. The lowest value comes from our own instrument, the only definition
that shares no cross-experiment input with $\Om$ and, by construction, the
noisiest of the five. The three definitions built from de-noising models,
which are the most committed to treating $\Om$ as pure contamination, all
return higher values ($0.666$ to $0.760$). We therefore report the most
conservative of the available answers as our headline. Under none of the five
definitions is $\Om$ rank-uninformative about true alignment.

\paragraph{A consistency check, not independent validation.} Read in the other
direction, the instrument agrees closely with the estimators themselves:
\texttt{tc\_rho} at $0.964$ ($n=63$), \texttt{pda} at $0.961$,
\texttt{re\_rho} at $0.899$ and \texttt{ols\_r} at $0.897$ ($n=69$ each), all
above our headline $0.651$. That is expected. \texttt{tc\_rho} and cross-half
agreement estimate the same parameter, the latent proxy--north-star
correlation, by different routes: one subtracts an estimate of $\Om$, the
other makes the sampling errors independent by construction. Their agreement
shows that the instrument measures what it should and that the corrections
estimate latent alignment well when given all 262 experiments; our claim is
that they rank worse at effective archive sizes like ours
(Section~\ref{sec:estvsrank}). These agreements cannot, however, validate the
estimators' ranking ability: estimators and instrument use the same effect
estimates, so their errors are dependent, and \texttt{tc\_rho} and
\texttt{pda} share a numerator. The stronger evidence is the $0.651$ of
Figure~\ref{fig:rankrank}: although computed on the same archive, the two
statistics plotted there share no sampling error (the halves are disjoint), no
cross-experiment fit and no subtraction.

\subsection{Ruling out estimation error as the explanation}
\label{sec:precision}

A remaining concern is that the association is an artifact of estimation error
in $\Om$. The archive rules this out directly. Across 200 random half-splits,
the mean within-experiment sampling covariance $\vcross$ reproduces across
halves with a median Pearson correlation of $0.9987$ (5th percentile
$0.9950$; median Spearman $0.9982$), and the sampling-error correlation at
$0.9930$ (5th percentile $0.9741$): the quantity the corrections
discard orders the candidates almost identically on independent halves. The
quantity needed to rescale a de-noised covariance to a correlation, the
de-noised north-star variance, is far less precise: the north star's
reliability has a bootstrap interval of $[-0.269,\,0.322]$ around $0.104$,
which includes zero. \texttt{tc\_rho} and \texttt{re\_rho} inherit that
imprecision, and Section~\ref{sec:tcrho} shows what it costs
\texttt{tc\_rho}. It does not explain why \texttt{tc\_slope} and \texttt{pda}
lose, since neither uses the north-star variance. Section~\ref{sec:mechanism}
shows that their cost arises because $\Om$ is estimated so accurately:
what they subtract is real, rank-aligned signal.

\subsection{Scope of the test}

On this archive, $\Om$ ranks the candidates in substantial agreement with each
of the five definitions of true alignment we consider. The test does not show
that $\Om$ equals $\Sig$, that de-noising is misconceived or that the
association holds on every archive. It shows that the premise justifying
de-noising for ranking is testable, with one additional pass over
customer-level data, and that it does not hold here. Where it holds, as in the
counterfactual world of Section~\ref{sec:ablation}, the corrections behave as
designed.

\section{The ranking cost of de-noising}
\label{sec:cost}

At this archive's size and reliability, removing $\Om$ costs ranking
accuracy; the cost grows with the dose removed, and it cannot be repaired by
estimating $\Om$ better. Section~\ref{sec:instrument} showed that $\Om$ is
rank-informative on this archive, so removing it discards information. This
section quantifies that cost, identifies its mechanism and shows that it is
not a blanket penalty on de-noising. The
archive-calibrated simulation, together with its counterfactual ablation
(Section~\ref{sec:ablation}), is the second of the paper's three main
analyses. The closed-form construction of Section~\ref{sec:mechanism} serves a
different purpose: it isolates the mechanism, showing why the cost
exists but not how large it is here.

\subsection{Why removing \texorpdfstring{$\Om$}{\textOmega12} can invert a ranking}
\label{sec:mechanism}

Consider the probability limits. As $K\to\infty$, an estimator that removes
$\Om$ exactly recovers a monotone function of $\Sig$, and its ranking becomes
exactly faithful. An estimator that keeps $\Om$ converges to a ranking on
$\Sig + \Om$, which is faithful when $\Om$ is constant across candidates or
increases with $\Sig$, but not in general. Asymptotically, then, the corrections
rank better, and our results confirm this.

At finite $K$ the picture changes, and the intuitive explanation is wrong. It
attributes the change to noise in $\widehat{\Om}$, which makes the correction
subtract the wrong amount. But $\widehat{\Om}$ is among the most reproducible quantities in the
archive (rank correlation $0.998$ across independent halves;
Section~\ref{sec:precision}), because it is estimated from millions of
customers per experiment, whereas the moment it is subtracted from rests on
$K$ experiments.

The mechanism is instead rank-aligned signal subtraction. The
corrected statistic a platform ranks on is a difference: a cross-experiment
moment, estimated from $K$ experiments and therefore noisy, minus a sampling
covariance, estimated from millions of customers and therefore very
precise. Subtracting a term whose own estimation variance is negligible leaves
the variance of the difference essentially unchanged while removing
between-candidate spread from it. Because $\Om$ is rank-aligned with $\Sig$ on
this archive (Section~\ref{sec:instrument}), the candidates it shrinks most are
the ones that should rank highest. As a result, the signal-to-noise ratio of
the ranking statistic falls, and it falls furthest where the ordering matters
most. As $K$ grows, the cross-experiment noise shrinks, the residual bias from
keeping $\Om$ becomes the binding constraint, and the ordering
reverses. This is why the result takes the form of a crossover in $K$ rather
than a general verdict on de-noising.

Three results separate this mechanism from the estimation-error explanation.
In the calibrated simulation, every correction is handed the exact
per-experiment sampling covariances rather than estimates of them
(Section~\ref{sec:sim}), so none can suffer from error in $\widehat{\Om}$,
yet those that remove $\Om$ still lose up to 40 points of pair accuracy to the
uncorrected baseline (Table~\ref{tab:dose2}); within a fixed reporting scale,
the loss grows monotonically with the amount subtracted
(Section~\ref{sec:dose}); and when $\Om$ is set to zero, so that the
corrections run unchanged but nothing rank-aligned remains to remove, the
ordering reverses (Section~\ref{sec:ablation}).

\begin{figure}[t]
\centering
\includegraphics[width=\textwidth]{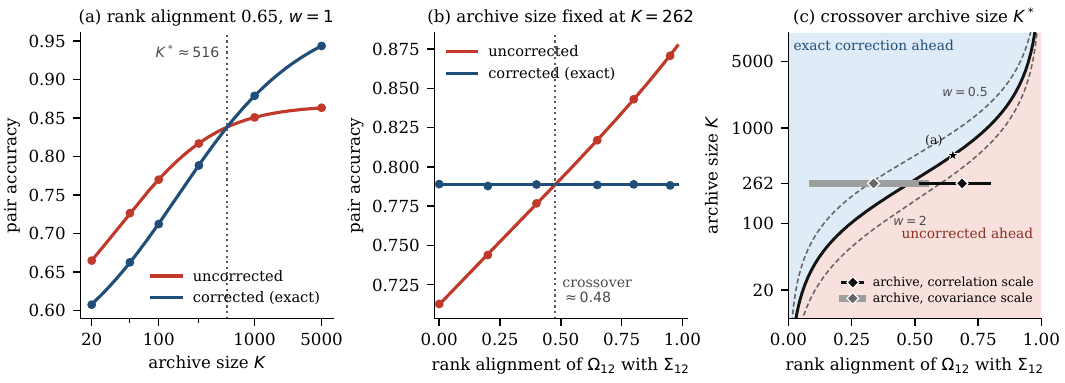}
\caption{\datasource{Stylized construction} The ranking cost persists when
$\Omega_{12}$ is subtracted exactly. \textbf{(a)}~Pair accuracy of the
uncorrected and corrected scores against $K$, at a chosen rank alignment of
$0.65$ and $w = 1$. \textbf{(b)}~Pair accuracy against rank alignment at
$K = 262$ and $w = 1$. \textbf{(c)}~The crossover size $K^{*}$ of
equation~\eqref{eq:crossover} (solid: $w = 1$; dashed: $w = 0.5$ and $2$);
the star is the crossover of panel (a). The bars at $K = 262$ place the real
archive's measured alignment on the correlation (black) and covariance (gray)
scales of Section~\ref{sec:instrument}: point estimate and 95\% candidate
bootstrap interval. Markers are Monte Carlo means over $4{,}000$ replicates;
lines are the closed form of equation~\eqref{eq:mechanism}, which matches
every marker to within $0.004$.}
\label{fig:mechanism}
\end{figure}

\paragraph{A construction without estimation error.} Figure~\ref{fig:mechanism}
checks the argument directly, by reducing the problem to its essentials. Each
candidate has a true cross-experiment moment
$\Sigma_{12}$ and a sampling-error moment $\Omega_{12}$ whose ranks agree with
those of $\Sigma_{12}$ at a controlled rank correlation; an archive of $K$
experiments measures their sum with noise. The uncorrected score is that
measurement. The corrected score subtracts $\Omega_{12}$ itself rather than an
estimate of it, so it equals the truth plus mean-zero noise and is unbiased
for the ranking target. No $\widehat{\Om}$ enters the construction.

The corrected score can still lose, and the reason can be seen in closed form.
Both scores carry the same measurement noise, which comes from having only $K$
experiments and is not removed by any subtraction; but the uncorrected score
retains a rank-aligned signal term that the corrected score has removed.
Because the construction is a Gaussian copula, both curves in
Figure~\ref{fig:mechanism} can be written exactly with two classical
identities for jointly normal variables \citep{Kruskal1958}: a score whose
correlation with the truth is $q$ orders a random pair correctly with
probability $\tfrac12 + \arcsin(q)/\pi$, and a Spearman correlation
$r_{\mathrm S}$ corresponds to a Pearson correlation of
$2\sin(\pi r_{\mathrm S}/6)$. With $g = c^2/K$ the
noise-to-signal ratio of the measurement, $r$ the Pearson correlation between
$\Sigma_{12}$ and $\Omega_{12}$ across candidates, obtained in this way from
their rank alignment, and $w$ the ratio of their
spreads,
\begin{equation}
\label{eq:mechanism}
\begin{aligned}
\mathrm{acc}_{\text{unc}}
  &= \tfrac12 + \tfrac{1}{\pi}\arcsin\!\frac{1 + wr}{\sqrt{1 + w^2 + 2wr + g}},\\[2pt]
\mathrm{acc}_{\text{cor}}
  &= \tfrac12 + \tfrac{1}{\pi}\arcsin\!\frac{1}{\sqrt{1 + g}} .
\end{aligned}
\end{equation}
The second expression contains neither $r$ nor $w$. Within this construction,
the mechanism is therefore an identity rather than a simulation result: once
$\Omega_{12}$ has been removed, the score cannot benefit from how informative
$\Omega_{12}$ was, however accurately it was removed. Panel (b) shows this
identity: the corrected curve is exactly flat in rank alignment, while the
uncorrected curve rises steadily with it. Equating the two expressions gives
the crossover archive size in closed form,
\begin{equation}
\label{eq:crossover}
K^{*}(r, w) \;=\; \frac{c^{2}\,r\,(2 + wr)}{w\,(1 - r^{2})},
\end{equation}
which is increasing in $r$: the more rank-aligned the sampling error, the
larger the archive must be before removing it pays off.

The construction does not, however, locate our archive on this map. Its
parameters are chosen, not measured: panel (a) fixes an alignment of $0.65$
(crossover $K^{*} \approx 516$); both (a) and (b) assume equal spreads
($w = 1$); and $c$ is set so that the uncorrected score reaches $0.817$ at
$K = 262$, the calibrated simulation's baseline. The archive's own alignment,
measured in Section~\ref{sec:instrument}, gives two readings whose
candidate-bootstrap intervals barely overlap: $0.686$ on the correlation scale
and $0.336$ on the covariance scale, which imply $K^{*} \approx 605$ (interval
$303$--$1{,}106$) and $151$ ($28$--$352$) at $w = 1$, on either side of our
$K = 262$. In panel (c), the first interval lies wholly where the uncorrected
score is ahead and the second straddles the crossover; halving or doubling $w$
moves the crossover alignment at $K = 262$ from $0.48$ to $0.32$ or $0.60$.
The quantitative answer must therefore come from the archive-calibrated
simulation (Sections~\ref{sec:dose} and~\ref{sec:map}) and the held-out
comparison (Section~\ref{sec:horserace}).

What the construction does establish is the qualitative claim it was designed
for, and that claim does not depend on the parameters: for any positive
alignment there is a range of archive sizes over which exact subtraction
loses, and the cost arises from removing a rank-aligned quantity, not
from removing it imprecisely. Better estimation of $\Om$ therefore cannot
reduce it. Whether the cost is worth paying is an empirical question in
$(\text{reliability},\,K)$, which we map next.

\subsection{Simulation design}
\label{sec:sim}

We build a data-generating process calibrated to the archive's own measured
moments (per-experiment sampling covariance matrices, the cross-experiment
effect covariance, sample sizes and the heavy tail of the empirical effect
distribution) under
two alternative calibration arms. The \emph{random-effects} arm fits the
cross-experiment covariance with a random-effects model; the
\emph{tail-correlation} arm fits it with the total-covariance estimator of
\citet{Bibaut2024}. The two arms thus calibrate the true correlations with
different estimators. We report both throughout, so that no result rests on a
calibration choice favorable to our conclusion. Appendix~\ref{app:dgp} gives
the full specification.

\paragraph{The corrections receive the exact $\Om$.} One design choice is
central to interpreting everything that follows. In each replicate, the
generator draws the sampling errors using a per-experiment covariance matrix
$\Omega_k$ and then passes those same matrices to every estimator. The
corrections do not estimate $\Om$, do not receive a perturbed version of it
and incur no plug-in error: they operate on the oracle. Whatever they lose in
the simulations that follow is therefore not the cost of estimating $\Om$
poorly but the cost of removing it correctly. On a real archive $\Om$
must be estimated, so the simulation is, if anything, favorable to the
corrections.

The archive plays three roles, which we keep separate. It supplies the moments
to which the generator is calibrated, and it fixes the size and effective size
at which we read the resulting map (Section~\ref{sec:map}); neither role is,
on its own, evidence for the conclusion. Only its third role, as the test bed
for the held-out comparison (Section~\ref{sec:horserace}), is an
out-of-sample test.

\subsection{The inversion}
\label{sec:inversion}

\begin{figure}[t]
\centering
\includegraphics[width=\textwidth]{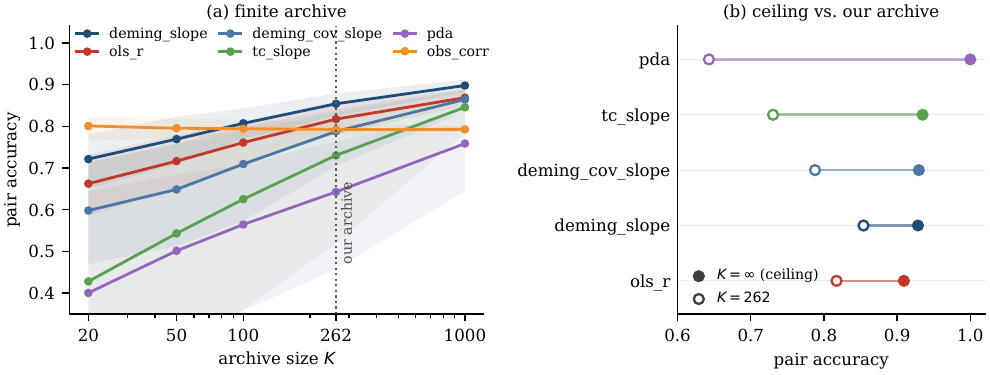}
\caption{\datasource{Calibrated simulation} The asymptotic and finite-sample
orderings are nearly reversed. Pair accuracy, simulated at the archive's
measured north-star reliability ($0.104$) in the random-effects arm, with the
sampling covariance as measured.
\textbf{(a)}~Accuracy against archive size $K$: replicate means, with the
5th--95th percentiles across replicates shaded. \textbf{(b)}~Each estimator's
analytic ceiling at $K = \infty$ (filled) against its accuracy at our
archive's $K = 262$ (open). \texttt{obs\_corr}, flat in $K$, appears only in
(a); \texttt{tc\_rho}, frequently undefined at these sizes
(Section~\ref{sec:tcrho}), in neither.}
\label{fig:accK}
\end{figure}

Figure~\ref{fig:accK} presents the paper's main quantitative result. Panel (b)
sets each estimator's asymptotic ceiling against its accuracy at our archive's
size, so the length of each row is what a finite archive costs. The ceilings
are ordered as theory predicts: \texttt{pda} is exactly rank-faithful in the
limit ($1.000$), followed by \texttt{tc\_slope} ($0.935$),
\texttt{deming\_cov\_slope} ($0.930$) and \texttt{deming\_slope} ($0.928$),
with the uncorrected baseline last at $0.909$; as expected, every correction
outperforms it asymptotically. At $K = 262$ the ordering is nearly reversed
(Spearman $-0.9$ between the two rankings): \texttt{deming\_slope} $0.854$,
uncorrected $0.817$, \texttt{deming\_cov\_slope} $0.788$, \texttt{tc\_slope}
$0.730$, \texttt{pda} $0.643$. Across the four corrections, both quantities
move monotonically with the dose of $\Om$ removed, in opposite directions: the
ceiling rises ($0.928$, $0.930$, $0.935$, $1.000$) and the gap below it widens
($0.074$, $0.142$, $0.204$, $0.357$). The larger a correction's asymptotic
ceiling, the further below it the correction falls at $K = 262$. Panel (a)
shows that the reversal is not specific to one archive size: across the whole
grid, \texttt{deming\_slope}, the correction with the lowest ceiling, leads
the other corrections and the uncorrected baseline, and \texttt{pda}, the one
with the highest, is last among the estimators shown. The purely
observational \texttt{obs\_corr} is asymptotically biased and so flat in $K$,
yet at $K = 20$ it ranks above every other estimator.

The inversion is a finite-sample effect, not an error in the asymptotic
calculation: in 95 of the 100 cells of our grid at $K = 262$, no estimator's
accuracy exceeds its own analytic ceiling. The five exceptions are Deming variants at
reliability $0.951$, far from our archive, where the ceilings are near or
below chance ($0.36$--$0.55$) and only approximate (Appendix~\ref{app:dgp}).

\subsection{The cost scales with the dose}
\label{sec:dose}

Table~\ref{tab:dose2} orders the estimators by how much of $\Om$ they remove
and reports pair accuracy at the archive's size. Within each reporting scale,
the relationship is monotone in the dose.

\begin{table}[t]
\centering
\small
\caption{\datasource{Calibrated simulation} Pair accuracy at $K=262$,
north-star reliability $0.104$,
random-effects arm, from the reliability sweep. Estimators are grouped by
reporting scale and ordered within each group by the dose of $\Om$ removed.}
\label{tab:dose2}
\begin{tabular}{@{}llc@{}}
\toprule
\textbf{Estimator} & \textbf{$\Om$ treatment} & \textbf{Pair acc.} \\
\midrule
\multicolumn{3}{@{}l}{\emph{Slope scale}} \\
\texttt{deming\_slope}      & kept; $x$-variance corrected  & $\mathbf{0.854}$ \\
\texttt{deming\_cov\_slope} & partially removed             & $0.788$ \\
\texttt{tc\_slope}          & removed from numerator        & $0.730$ \\
\addlinespace
\multicolumn{3}{@{}l}{\emph{Correlation and score scales}} \\
\texttt{ols\_r}             & kept (uncorrected baseline)   & $0.817$ \\
\texttt{pda}                & removed; $\div\,\hat\sigma_x$ & $0.643$ \\
\texttt{tc\_rho}            & same dose; also $\div\,\hat\sigma_y$ & $0.421$ \\
\addlinespace
\multicolumn{3}{@{}l}{\emph{Reference statistic}} \\
\texttt{obs\_corr}          & \emph{is} $\Om$               & $0.793$ \\
\bottomrule
\end{tabular}
\end{table}

We group by scale because the reporting transformation is a second axis, and
conflating it with the dose would misattribute the losses: a slope and a
correlation formed from the same de-noised covariance can rank differently,
because the correlation divides by de-noised variances that must themselves be
estimated. Holding the scale fixed isolates the dose. On the slope scale,
accuracy falls $0.854 \to 0.788 \to 0.730$ as more of $\Om$ is removed; on the
correlation and score scales, it falls from $0.817$ to $0.643$ when $\Om$ is
removed.

None of these gaps is within the simulation's own noise. On the same grid and
replicates, the paired bootstrap 95\% interval for each estimator's gap to the
uncorrected baseline is $[+0.035,\,+0.039]$ for \texttt{deming\_slope},
$[-0.033,\,-0.026]$ for \texttt{deming\_cov\_slope}, $[-0.095,\,-0.079]$ for
\texttt{tc\_slope}, $[-0.182,\,-0.167]$ for \texttt{pda} and
$[-0.427,\,-0.368]$ for \texttt{tc\_rho}; every one excludes zero. What these
intervals pin down is the expectation, not the outcome on any one
archive. A single archive of $262$ experiments is far noisier: across
replicates, the 5th-to-95th percentile range of \texttt{tc\_rho}'s pair
accuracy runs from $0.00$ to $0.75$, against $0.78$ to $0.85$ for the
uncorrected baseline. The dose ladder describes what a platform should expect
on average; it is not a guarantee about any particular archive.

\texttt{pda} and \texttt{tc\_rho} are not two rungs of the dose ladder. They
share a numerator and differ only by $\hat\sigma_y$, a candidate-independent
north-star factor, so on common support they have the same ranking
limit and the same analytic ceiling of $1.000$. The $22.2$-point gap between
them is therefore not an additional dose of de-noising but the finite-sample
cost of one more estimated variance in the denominator, which can come out
negative and leave the estimator undefined (Section~\ref{sec:tcrho}). Their
nearly identical held-out rewards ($0.124$ and $0.126$) are a consistency
check on this reading rather than two independent observations.

The best estimator in the table is itself a correction (a Deming
regression that corrects the proxy-side variance while leaving $\Om$ intact),
and it outperforms the uncorrected baseline by 3.7 points. Our finding is
therefore not that corrections fail, but that the ranking cost scales with how
much of the sampling-error covariance is removed, and that the losses are
concentrated at the aggressive end of that scale.

\texttt{obs\_corr} is listed separately because it is not a point on the
gradient but its limiting case: the within-experiment sampling-error
association alone, which uses no cross-experiment variation. That it
reaches $0.793$, ahead of estimators that use all 262 experiments to remove
this very quantity, illustrates how much rank information $\Om$ carries on
this archive. We do not propose it as an estimator: it has no asymptotic
justification and none of the properties a platform would want. We report it
as a reference point.

\subsection{The mechanism, identified by ablation}
\label{sec:ablation}

\begin{figure}[t]
\centering
\includegraphics[width=0.63\textwidth]{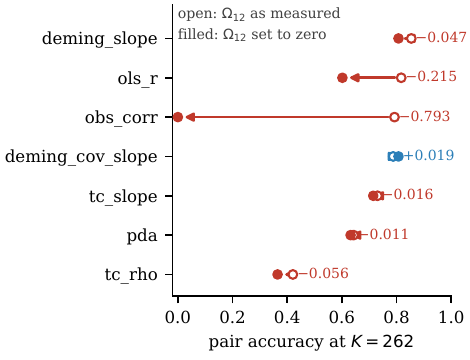}
\caption{\datasource{Calibrated simulation} Counterfactual ablation of the
within-experiment sampling-error covariance. Pair accuracy at $K = 262$ and
the archive's north-star reliability, in the simulated world as measured (open
circle) and in the same world rebuilt with $\Omega_{12}$ set to zero (filled
circle); random-effects arm. Arrows and numbers give the change: red where
accuracy falls, blue where it rises.}
\label{fig:ablation}
\end{figure}

A natural concern about Figure~\ref{fig:accK} is that the simulation is built
in a way that favors the uncorrected baseline. Figure~\ref{fig:ablation}
addresses it with a counterfactual: we rebuild the same simulated world with
$\Om$ set to zero and rerun every estimator.

If the baseline's lead came from an artifact of the simulation harness, it
would persist. Instead, the baseline falls from $0.817$ to $0.602$, a
drop of $21.5$ points that moves it from second of seven to fifth and leaves
it behind every correction except \texttt{tc\_rho}. Meanwhile,
\texttt{tc\_slope} ($0.730 \to 0.715$) and \texttt{pda} ($0.643 \to 0.632$)
barely move, as expected of estimators that discard $\Om$ when $\Om$ is
absent.

Two further results confirm that the ablation acts on the mechanism rather
than on the harness. First, \texttt{deming\_cov\_slope} is the only
estimator that improves ($+1.9$ points), and its value coincides with
\texttt{deming\_slope}'s ($0.8072$ for both, to four decimals), as it should:
with $\Om = 0$, the covariance-aware Deming estimator reduces to plain Deming.
The full simulation pipeline thus recovers an exact algebraic identity
numerically. Second, \texttt{obs\_corr}, which is $\Om$ expressed as a
correlation, falls to $0.000$: with $\Om$ removed, the statistic is
identically zero across candidates, so it orders no pair correctly, and its
unconditional pair accuracy is zero rather than the $0.5$ of a coin flip.

The interpretation therefore cuts both ways: the corrections behave as
designed when their premise holds and lose ranking accuracy when it does
not. Which of the two worlds a platform is in can be tested with the
instrument of Section~\ref{sec:instrument}.

\subsection{Ranking loss is not estimation loss}
\label{sec:estvsrank}

\texttt{re\_rho}, the random-effects estimator of the cross-experiment
true-effect correlation, illustrates most clearly that the choice of loss
drives the result. Judged as an estimator of $\rho$, the quantity the
platform says it wants, it is the only one in the study that is essentially
unbiased at the archive's size: its bias shrinks from $-0.226$ at $K = 20$
through $-0.022$ at $K = 100$ to $+0.006$ at $K = 262$, whereas the
uncorrected baseline's stays near $-0.187$ at every $K$, and
\texttt{tc\_rho}'s is still $-0.062$ at $K = 262$ on the estimates it
returns.

It does not, however, win the estimation problem outright: removing the bias
costs more variance than the bias was worth (RMSE $0.263$ against the
baseline's $0.216$ at $K = 262$, and $0.826$ against $0.350$ at $K = 20$,
where its bias advantage has also gone), so the variance cost that concerns
this paper appears within the estimation problem as well. We make the
ranking argument at $K = 262$ because, of the sizes we simulate, that is where
the estimation case for the correction is strongest.

Even there it ranks worse: its pair accuracy is $0.721$ against the
baseline's $0.819$, about ten points lower, and its top-5 recovery is $0.347$
against $0.667$, so it identifies barely half as many of the truly best
candidates. An estimator can be consistent, and essentially unbiased at the
available sample size, and still put the wrong metric at the top of the
shortlist. There is no paradox here: a bias common to all candidates does not
affect a ranking, whereas estimation noise specific to each candidate does. It
does, however, caution against validating proxy-selection methods on
estimation criteria.

\subsection{Three loss functions and what they cannot settle}
\label{sec:threelosses}

All three losses put full de-noising behind the uncorrected baseline
(Table~\ref{tab:losses}). They part ways over whether \texttt{deming\_slope}
or the baseline itself ranks best, and the answer depends on which part of
the ranking a platform uses.

\begin{table}[t]
\centering
\small
\caption{\datasource{Calibrated simulation} Three losses at $K = 262$,
random-effects arm, from the primary simulation. In parentheses, for the
estimators that leave some pairs unresolved: pair accuracy over the resolved
pairs only (Section~\ref{sec:tcrho}).}
\label{tab:losses}
\begin{tabular}{@{}lccc@{}}
\toprule
\textbf{Estimator} & \textbf{Pair acc.} & \textbf{Top-5 regret} & \textbf{Top-5 rec.} \\
\midrule
\texttt{deming\_slope}      & $\mathbf{0.853}$ & $0.0414$ & $0.523$ \\
\texttt{ols\_r}             & $0.819$ & $\mathbf{0.0123}$ & $\mathbf{0.667}$ \\
\texttt{deming\_cov\_slope} & $0.791$ & $0.0629$ & $0.408$ \\
\texttt{tc\_slope}          & $0.737$\;($0.748$) & $0.0862$ & $0.341$ \\
\texttt{re\_rho}            & $0.721$ & $0.0743$ & $0.347$ \\
\texttt{pda}                & $0.648$\;($0.658$) & $0.1331$ & $0.211$ \\
\texttt{tc\_rho}            & $0.427$\;($0.676$) & $0.1675$ & $0.176$ \\
\bottomrule
\end{tabular}
\end{table}

\paragraph{The whole ranking.} On pair accuracy, the mild correction
\texttt{deming\_slope} leads and the uncorrected baseline is second.

\paragraph{The head of the ranking.} This lead does not carry over to the head
of the list: on top-5 regret, the uncorrected baseline is best at $0.0123$, and
\texttt{deming\_slope} is $3.4$ times worse at $0.0414$; on top-5 recovery, the
baseline finds $0.667$ of the true top five against \texttt{deming\_slope}'s
$0.523$. Both head-of-list losses agree that the aggressive corrections are
the most costly: \texttt{pda} and \texttt{tc\_rho} incur an order of magnitude
more regret than the baseline and recover only about a fifth of the true top
five. A
platform that shortlists five candidates should read the two right-hand
columns; a platform that ranks a long tail should read the left.

\paragraph{Where the errors fall.} The losses disagree because the estimators
misplace candidates in different parts of the ranking. Computing each
estimator's mean rank displacement (the absolute distance between a
candidate's measured and true rank, as a fraction of the catalog's length)
within each tenth of the true ordering separates them clearly. The
uncorrected baseline's errors sit in the middle of the ranking, the part a
shortlisting platform does not use, and it becomes far more accurate at the
head of the list. The aggressive corrections improve much less there, and
\texttt{pda} even displaces candidates more severely at the head than in the
middle, so at the head of the list they misplace candidates several times more
than the baseline does. Whole-ranking pair accuracy averages over a region
where every estimator struggles, whereas the head of the list is where
retaining $\Om$ helps most. The tail-correlation arm shows the same contrast.

\paragraph{Both calibration arms agree.} The second calibration arm supports
this reading. In the tail-correlation arm, the uncorrected baseline is again
best on both head-of-list losses (top-5 regret $0.021$ against
\texttt{deming\_slope}'s $0.060$; top-5 recovery $0.605$ against $0.553$), and
the dose ordering of Figure~\ref{fig:accK} is reproduced: accuracy again falls
monotonically as more of $\Om$ is removed ($0.829$, $0.821$, $0.765$ and
$0.708$ for \texttt{deming\_slope}, \texttt{deming\_cov\_slope},
\texttt{tc\_slope} and \texttt{pda}), and \texttt{tc\_rho} is again last at
$0.416$. One comparison, at the mild end of the ladder, does not replicate:
\texttt{deming\_cov\_slope} is $2.8$ points behind the baseline in the random-effects arm and $0.8$ points ahead of it in the
tail-correlation arm, so the direction of that small gap remains open. At the aggressive
end (\texttt{tc\_slope}, \texttt{pda} and \texttt{tc\_rho}), which carries our
main claims, the two arms agree on both sign and ordering, with margins of
$4.8$ to $39.7$ points.

\subsection{Does the simulation resemble the archive?}
\label{sec:calibration}

The simulated accuracies describe the real archive only as far as the
generator reproduces it. Appendix~\ref{app:calib} tests this by placing each
statistic of the real archive within its distribution across simulated
archives, and two of its results bear on the conclusions above.

For \texttt{ols\_r}, the baseline against which every correction is judged,
the archive's realized departure from the simulated median is small in both
calibration arms, and no individual candidate deviates by more than $1.6$
simulated standard deviations. The generator shows no systematic error on the
quantity the paper relies on most.

One statistic is clearly mis-calibrated in both arms: the share of
experiments in which a candidate reaches significance. The generator
reproduces it on average but not candidate by candidate, because its heavy
tail is calibrated to the archive as a whole rather than to each candidate
(Appendix~\ref{app:calib}). Since we use \texttt{share\_significant} only as
measured on the real archive, this failure does not affect our conclusions.

The audit cannot rule out one possibility that would flatter our results:
the generator may inject more replicate-to-replicate noise than the archive
contains. If it does, every simulated accuracy is pessimistic and, because
extra noise acts like a smaller archive (Section~\ref{sec:mechanism}), the
simulated crossover lies at larger archives than the true one, overstating
how far below it our archive sits. The evidence that does not depend on the
generator, the held-out comparison of Section~\ref{sec:horserace}, is therefore the
important check; it reproduces the simulated ordering on real experiments.

\subsection{Why \texttt{tc\_rho} ranks last: it is often undefined}
\label{sec:tcrho}

\texttt{tc\_rho} is last in every table above, but its position says more
about how often it fails to return a usable score than about how well it
orders the candidates it does score. The estimator requires a variance estimate that can be
negative, in which case it is undefined; and even when the variance is
positive, the resulting ratio can fall outside $[-1, 1]$, in which case it is
inadmissible as a correlation. We count both failures together.
\texttt{tc\_rho} is undefined or inadmissible for 6 of 69 candidates at the
point estimate and in $42.5\%$ of bootstrap resamples, mostly because the
de-noised variance is negative. At the point
estimate all six are finite ratios above one, and they are the six candidates
that \texttt{pda} ranks highest: \texttt{tc\_rho} is \texttt{pda} divided by
the common factor $\hat\sigma_y$, so the strongest candidates are the first
to leave the unit interval. In simulation, \texttt{tc\_rho} resolves only
$19\%$ of candidate pairs at $K = 20$, rising to $62\%$ at $K = 262$; and in
the published DGP of \citet{Bibaut2024} itself, whose true-effect correlation
is $-0.995$, it is undefined or inadmissible in $49\%$ of draws at
$n = 5{,}000$ customers per arm.

Scored only where it is defined, its pair accuracy at $K = 20$ is $0.578$,
comparable to its peers, against an unconditional $0.115$; at $K = 262$ the
same contrast is $0.676$ against $0.427$ (Table~\ref{tab:losses}). The
accurate summary is therefore not that this estimator ranks badly, but that at
realistic archive sizes it is frequently undefined, and a platform that adopts
it needs a rule for handling those cases, which can include its most promising
candidates. We regard this as a property of the correlation scale rather than
of the estimator: rescaling a de-noised
covariance by de-noised variances requires all three to be well behaved, and at
$K$ in the tens they are not.

\section{Locating real archives on the map}
\label{sec:map}

Our archive lies on the side of the crossover where the aggressive
corrections lose, as would published archives of similar size that shared its
structure and north-star reliability; an out-of-sample test on our archive
reproduces the predicted ordering. The results so far are conditional on a
regime. This section locates real archives, ours and others', within it and
then tests the prediction on held-out experiments. The held-out
comparison of Section~\ref{sec:horserace} is the third of the main analyses,
and the only one scored on realized decisions rather than on a simulated or
latent truth.

\subsection{The crossover map}

\begin{figure}[t]
\centering
\includegraphics[width=0.63\textwidth]{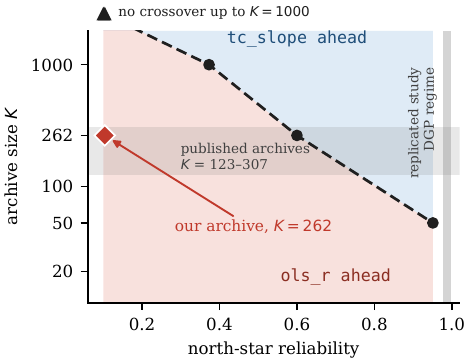}
\caption{\datasource{Calibrated simulation} Where correcting starts to pay.
For each north-star reliability set in the simulation, the dashed curve
gives the smallest archive size $K$ at which
\texttt{tc\_slope} overtakes the uncorrected baseline \texttt{ols\_r} on pair
accuracy; random-effects arm, sampling covariance as measured. The shaded
regions refer to this pair only. Each dot is the first of five grid sizes at
which \texttt{tc\_slope} leads (paired 95\% interval of its gap above zero),
so the crossing lies between it and the next smaller grid size. Horizontal
band: the three published archives whose experiment counts we could verify,
$123$ \citep{Chou2025}, $200$
\citep{Zhang2024} and $307$ \citep{Tripuraneni2024}, placed on the size axis
only because none reports north-star reliability. Vertical band: the
north-star reliability of the simulation design of \citet{Bibaut2024}. Diamond:
our archive at its nominal size (Section~\ref{sec:neff}).}
\label{fig:crossover}
\end{figure}

Figure~\ref{fig:crossover} maps the crossover in the random-effects arm. We
count an estimator as leading at a grid size only when the paired 95\% interval
of its gap to \texttt{ols\_r} (Section~\ref{sec:dose}) lies above zero. At
our archive's measured north-star reliability of $0.104$, \texttt{tc\_slope}
never overtakes the uncorrected baseline anywhere on our grid, whose largest
size is $K = 1000$;
the same holds at reliability $0.10$. A crossing appears only when reliability
is raised well above anything our archive offers: \texttt{tc\_slope} first
leads at $K = 1000$ at reliability $0.373$, at $K = 262$ at $0.60$ and at
$K = 50$ at $0.951$ (upper bounds on the crossing, given the grid). The more
aggressive estimators fare worse still. \texttt{pda} and \texttt{tc\_rho}
never overtake the baseline at any simulated size except at reliability
$0.951$, where they need $K = 262$ and $K = 1000$, respectively. At the other
end, \texttt{deming\_slope}, the mildest correction, is ahead at $K = 20$ at
every reliability rung, consistent with Section~\ref{sec:dose}. The
tail-correlation arm agrees at our archive's reliability, where none of
\texttt{tc\_slope}, \texttt{pda} and \texttt{tc\_rho} leads at any size, but
moves the thresholds at higher reliabilities in both directions. In that arm,
\texttt{tc\_slope} leads only at reliability $0.951$, from $K = 100$;
\texttt{tc\_rho} never leads; and \texttt{deming\_slope} trails at every size
at $0.951$. \texttt{pda}, by contrast, already leads at $K = 1000$ at
reliability $0.60$, by $0.005$, after trailing by $0.024$ at $K = 262$.

Two bands complete the figure. The horizontal band marks the three published
archives whose size we could verify against a primary source; larger counts
circulate, but we could not verify them as counts of distinct real
experiments. For every archive in the band, then, and at every
reliability rung we simulate up to and including $0.60$, not correcting on the
correlation scale ranks better than correcting with \texttt{pda} or
\texttt{tc\_rho}.

Placing those archives on the map holds four things fixed at our archive's
measured values: its rank alignment, its spread ratio $w$, its
per-experiment precision profile and its candidate catalog. The first of these
matters a great deal: read at equal spreads, the two measurements of our own
archive's rank alignment imply crossover sizes on either side of our own $K$
(equation~\eqref{eq:crossover}, Section~\ref{sec:mechanism}). None of the three
papers reports a rank alignment, a spread ratio or a north-star reliability,
and we impute none of them. The band therefore says where the crossover lies
under our archive's structure, and how far published sizes sit from it; it is
not a claim about those studies, whose structural parameters we do not know.
Our own archive is plotted at its nominal $K = 262$, which
Section~\ref{sec:neff} shows is conservative; its reliability rung of
$0.104$, slightly below the generator's implied $0.125$, errs the other way,
placing us marginally further below the crossover than the generator's own
calibration would (Appendix~\ref{app:dgp}). The vertical band marks the
$0.976$--$0.999$ regime of the published simulation design, where the
corrections work as designed and which no metric in our archive occupies
(Figure~\ref{fig:ladder}).

\subsection{Effective archive size}
\label{sec:neff}

$K = 262$ overstates our archive. Because experiments differ greatly in
precision, a handful of them dominate the fits. We measure this by the
\emph{effective leverage count on the proxy axis}: the inverse Herfindahl
index of each experiment's squared deviation from the mean proxy effect,
$\neff = (\sum_k d_k^2)^2 / \sum_k d_k^4$ with
$d_k = \hat\tau^x_k - \bar{\hat\tau}^x$, which is the quantity that governs how
concentrated the cross-experiment slope is. It is Kish's effective-sample-size
formula \citep{Kish1965} applied to leverage rather than to survey weights, and
it is the appropriate statistic here because the cross-experiment fit is an unweighted ordinary
least-squares regression (Appendix~\ref{app:estimators}), in which an
experiment's weight in the slope is its squared deviation $d_k^2$, not its
precision. Its median across candidates is $10.0$ (mean $13.8$, minimum $2.4$,
maximum $36.4$). All 69 candidates fall below $\neff = 50$, against a nominal
$K = 262$. The same two experiments are the single highest-leverage
observation for 88\% of candidates (one for 47 of 69, the other for 14).
Closed-form standard errors understate bootstrap standard errors by a median
factor of $1.81$ (maximum $2.85$), so the nominal precision of these fits is
itself optimistic.

This is not a second discount to apply to Figure~\ref{fig:crossover}, because
the figure's size axis already incorporates it: the simulation draws each
experiment's precision from the archive's own templates, so a simulated
archive of nominal size $262$ has a median $\neff$ of $13.1$, against $10.0$
for the real one. The small residual gap runs in the conservative direction,
since the real archive is slightly weaker than the simulated one, so a
platform can locate itself on the figure by nominal size. What $\neff$
adds is an explanation of why 262 experiments sit so far below the crossover
(in leverage terms, they are worth roughly a dozen) and one caveat: a platform
whose archive is more evenly balanced than ours has a higher $\neff$ at the
same nominal $K$, so for it the curve errs on the side of not correcting, and
it would reach its crossover at a smaller nominal size than the figure
implies.

\paragraph{Influential experiments.} Leverage this concentrated means that the
two dominant experiments could, in principle, be driving our conclusions. They
are not. We deleted each of the 262 experiments in turn, then the two
dominant ones together, and after each deletion recomputed $\Om$, true
alignment under the four definitions of Table~\ref{tab:truth} other than
cross-half agreement, and the reliabilities that calibrate the simulation
(Appendix~\ref{app:influence}, Table~\ref{tab:influence}). Without the two
dominant experiments, the association between $\Om$ and true alignment
strengthens under all four definitions, to between $0.77$ and $0.93$. No single
deletion lowers it by more than $0.10$ except under the total-covariance
definition, which fails for the reason given in Section~\ref{sec:tcrho}: two
deletions leave so little north-star signal that the estimator falls outside
$[-1, 1]$ for half the candidates or more.

Single deletions cannot detect influence that several experiments exert only
jointly; replication on disjoint parts of the archive is the more demanding
test. Table~\ref{tab:designs} applies it to our instrument, which needs
customer-level data and is not refitted here. The instrument holds in both
halves and both domains, including the first half ($0.709$) and the
smaller domain ($0.907$), both of which exclude the two dominant
experiments. The deletions also leave the archive far from the crossover of
Figure~\ref{fig:crossover}: the north star's reliability stays at or below
$0.130$ throughout, while at $K = 262$ \texttt{tc\_slope} does not overtake the
baseline even at a reliability of $0.373$. Without the two dominant
experiments the median $\neff$ more than doubles, to $22.6$, which makes the
archive more balanced than a simulated one of the same size, the case the
caveat above concerns. Even placed by its effective size, however, it stays
below the crossover at our archive's reliability
(Appendix~\ref{app:influence}).

At nominal $K = 262$ and reliability $0.104$, the simulation makes a clear
prediction: the uncorrected and mildly corrected measures should rank best on
the real archive. The archive's own experiments can test this prediction
directly, without the generator.

\subsection{A contamination-free held-out comparison}
\label{sec:horserace}

\paragraph{Contamination in the evaluation criterion.} A naive retrospective
evaluation (rank candidates on some experiments, then score the winner against
the north star on the same experiments) is itself contaminated by $\Om$.
\citet{Chou2025} derive this bias analytically and propose a within-experiment
fold split, which we adopt. Our payoff functional differs from theirs (they
target a rule's cumulative return, we target the mean north-star effect per
firing event; Section~\ref{sec:limits} discusses what that difference
implies), but the fold split removes the same selection channel in both cases.
Our contribution here is not the correction itself but a measurement of its
magnitude on a real archive.

A zero-effect null isolates the bias. Hold the archive's per-experiment
sampling covariance matrices fixed, set every true effect to zero and reward
each decision of a significance rule on the proxy with the north-star estimate
from the same customers. The true reward is zero, but the expected measured
reward is $\rho_k \lambda(1.96)$ north-star standard errors, where $\rho_k$ is
the candidate's within-experiment sampling-error correlation and
$\lambda(1.96) = 2.338$ the inverse Mills ratio at the two-sided 5\%
threshold: a median of $+0.83$ across the 69 candidates,
positive for every one. On disjoint customer folds the same quantity is
$+0.0008$, within Monte Carlo error of zero, so the bias comes from the fold
overlap, not from the rule.

On the real archive, contamination inflates every measure of alignment, by
$+0.283$ in Spearman correlation on average, and inflates the uncorrected
baseline's reward by a factor of $2.26$ ($0.562$ same-sample against $0.249$
on disjoint folds). The one statistic with no alignment content,
\texttt{share\_significant}, moves the other way ($-0.13$): a channel that
inflates alignment and deflates mere significance behaves like selection on
noise, not like a mechanical artifact of our harness. We report both columns,
labeling the contaminated one as a positive control.

\paragraph{Design.} We split the archive by experiment into a ranking set and a
disjoint scoring set. Each candidate is scored by the measure under test fitted
on the ranking set, and rewarded by the average north-star effect per shipped
decision on the scoring set, with the trigger and the payoff read on
non-overlapping customer folds. A measure's reward is the Spearman correlation
across candidates between the two; Appendix~\ref{app:holdout} states the
protocol in full. Our headline test is a single cell of this design: the random experiment-level split at ranking size $131$, half the
archive. One hundred splits were attempted and $67$ produced a finite score for
every measure, at a mean of $58.0$ candidates scored per split. In the other
$33$, \texttt{tc\_rho} fails for so many
candidates (Section~\ref{sec:tcrho}) that too few remain to compare; we exclude
these splits rather than charge them to any measure. The
other ranking sizes ($20$, $50$, $100$) are reported as a learning curve, and
the temporal and two domain schemes as transfer tests in Appendix~\ref{app:domain};
some of those cells score too few splits to be informative, and we mark them
instead of averaging them in.

\begin{figure}[t]
\centering
\includegraphics[width=0.63\textwidth]{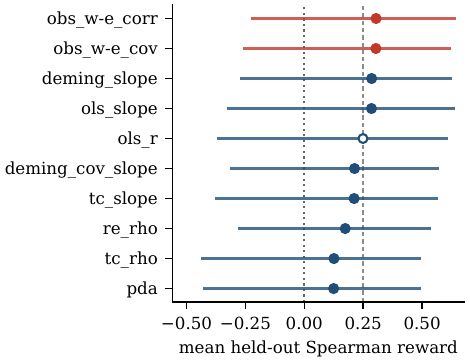}
\caption{\datasource{Real archive} Held-out reward over 67
contamination-free splits: the Spearman correlation across candidates between
each measure, fitted on one half of the experiments, and the candidates'
realized rewards on the other half (Section~\ref{sec:horserace}). Points:
means over splits; bars:
5th--95th percentiles. Open marker and dashed line: the uncorrected baseline
\texttt{ols\_r}. Red: the two purely observational measures. Labels abbreviate
\texttt{obs\_within\_experiment\_*} as \texttt{obs\_w-e\_*}.}
\label{fig:holdout}
\end{figure}

\paragraph{Result.} Figure~\ref{fig:holdout} reports the result, and the
ordering is broadly the one Section~\ref{sec:cost} predicts. The two purely
observational measures lead, at $0.305$ and $0.304$. Two fits that keep
$\Om$ intact follow and are indistinguishable from one another (the mildest
correction, \texttt{deming\_slope}, at $0.286$ and the uncorrected OLS slope
at $0.285$), and the correlation-scale uncorrected baseline \texttt{ols\_r}
comes next at $0.249$. The estimators that remove part or all of $\Om$
trail, from \texttt{deming\_cov\_slope} at $0.213$ and
\texttt{tc\_slope} at $0.211$ down to \texttt{tc\_rho} at $0.126$ and
\texttt{pda} at $0.124$, with \texttt{re\_rho}, the one estimator that the
simulation finds essentially unbiased for $\rho$, between them at $0.174$. The
mean rewards thus reproduce the dose gradient of
Section~\ref{sec:dose}. The marginal intervals in Figure~\ref{fig:holdout}
overlap heavily, because the archive's effective size is small, but the
paired comparison within splits is much more one-sided: across the $67$
scored splits, the share in which \texttt{pda} beats the uncorrected baseline
is $0.10$, against $0.12$ for \texttt{tc\_rho}, $0.30$ for \texttt{tc\_slope},
$0.70$ for the uncorrected slope and $0.61$--$0.66$ for the two observational
measures. The separation into three groups (the two observational measures,
the uncorrected slope and \texttt{deming\_slope} at the top; \texttt{ols\_r}
in the middle; the estimators that remove part or all of $\Om$ at the bottom)
holds at all four random-split ranking sizes, including the smallest, at which
a ranking set of $20$ experiments excludes both experiments that dominate the
fits of Section~\ref{sec:neff} with probability $0.85$. The ordering within
the top group does not (at size $100$, \texttt{deming\_slope} edges ahead;
Appendix~\ref{app:holdout}), but at every random-split size both
observational measures beat \texttt{ols\_r} on mean reward, although
\texttt{ols\_r} still leads in $29$--$48\%$ of individual splits. In every
random-split comparison, an in-sample convex blend of an observational measure
and \texttt{ols\_r} puts all its weight on the observational one, the
component the corrections discard (Appendix~\ref{app:holdout}).

\begin{figure}[t]
\centering
\includegraphics[width=0.67\textwidth]{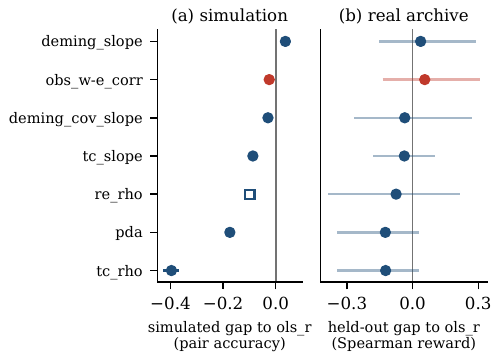}
\caption{\datasource{Simulation vs.\ real archive} Does the simulation predict
the real archive? Each row is one measure's gap to the uncorrected baseline
\texttt{ols\_r}; positive means ahead of it. \textbf{(a)}~Calibrated
simulation at $K = 262$ and the archive's north-star reliability
(random-effects arm): gap in pair accuracy, with 95\% intervals over
replicates, mostly narrower than the marker. \textbf{(b)}~Real archive: gap in
held-out Spearman reward (Figure~\ref{fig:holdout}), with 5th--95th
percentiles across the 67 splits. The units differ, so only signs and
orderings are comparable; rows are sorted by the simulated gap. Red: the one
measure whose sign disagrees. Open square: \texttt{re\_rho}, which the
reliability sweep omits, taken from the primary simulation at the same size
and arm. Rank correlation of the two sets of gaps: $0.93$ ($p = 0.003$
against a random relabeling of the seven measures).}
\label{fig:simreal}
\end{figure}

No single gap is significant on one archive: for every gap against the
uncorrected baseline, both the 5th--95th percentile range across splits and the
90\% bootstrap interval over candidates include zero (the tightest range among
the corrections, \texttt{tc\_slope}'s, is $[-0.179,\,+0.103]$), and we do
not convert the split-level shares above into $p$-values, because the splits
share experiments. With $\neff$ in the tens, the evidence has to lie in the
joint pattern, which Figure~\ref{fig:simreal} sets against the prediction of a
simulation calibrated before the comparison was run. The simulated and real
gaps are in different units, so we compare
only their signs and orderings. Treating each measure's gap as one comparison,
six of the seven have the sign the simulation predicts, and the correspondence
goes beyond the signs: the rank correlation between the simulated and the real
gaps across the seven measures is $0.93$, which a random relabeling of the
measures would produce with probability $0.003$. The exception is
\texttt{obs\_within\_experiment\_corr}, and the disagreement strengthens
rather than weakens our argument: the simulation places the pure
within-experiment association $2.5$ points of pair accuracy behind the
uncorrected baseline, whereas the real archive places it $5.6$ points of
held-out reward ahead. The two orderings differ only in two neighboring
pairs, one at each end. At the top, \texttt{obs\_within\_experiment\_corr}
and \texttt{deming\_slope} trade places. At the bottom, \texttt{tc\_rho} and
\texttt{pda} do, because the simulation counts \texttt{tc\_rho}'s missing
scores against it ($38\%$ of pairs) and the real comparison, which scores it
only where it is defined, does not (Section~\ref{sec:tcrho}); on the real
archive, the two penalties are level.

\begin{table}[t]
\centering
\small
\caption{\datasource{Real archive and simulation} Robustness checks. Each
tests whether the paper's conclusion could be an artifact of implementation,
calibration or evaluation design.}
\label{tab:controls}
\begin{tabular}{@{}P{0.30\textwidth}P{0.66\textwidth}@{}}
\toprule
\textbf{Check} & \textbf{Result} \\
\midrule
Set $\Om$ to zero (ablation) & Baseline falls from 2nd to 5th of 7
($0.817 \to 0.602$), behind every correction except \texttt{tc\_rho}
(Section~\ref{sec:ablation}) \\[2pt]
Invariance under the ablation & \texttt{tc\_slope} and \texttt{pda} move by
less than 2 points, as designed; \texttt{deming\_cov\_slope} coincides with
\texttt{deming\_slope}, as the algebra requires \\[2pt]
Subtract $\Omega_{12}$ exactly, without estimation error &
Cost persists at every positive rank alignment; at the construction's reference
alignment the crossover is $K^{*}\approx516$, and the archive's two measured
alignments imply crossovers on either side of $K=262$
(Figure~\ref{fig:mechanism}) \\[2pt]
Randomization integrity & No more sample-ratio flags than chance: after the
exclusion at $p < 0.001$, $2$ of the $262$ experiments fall below $p = 0.01$,
against $2.4$ expected, and the $p$-values are consistent with uniform
(Kolmogorov--Smirnov $p = 0.66$). No differential attrition: in no experiment
does the share of customers with complete metric records differ between the
arms by more than $0.38$ percentage points (Section~\ref{sec:archive}) \\[2pt]
Two calibration arms & The dose ordering replicates, and the baseline is best
on both head-of-list losses in both arms. Five of six gap signs against the
uncorrected baseline agree across the arms; the exception is the mild
correction \texttt{deming\_cov\_slope}, whose gap is small in both
($-0.028$ vs $+0.008$; Section~\ref{sec:threelosses}) \\[2pt]
Finite-sample accuracy vs analytic ceiling & At $K = 262$, 95 of 100 cells at
or below their analytic ceiling; the 5 exceptions are all Deming cells at
reliability $0.951$ (Section~\ref{sec:inversion}) \\[2pt]
Calibration audit against the real archive & \texttt{ols\_r} realized departure
$+0.08$ and $-0.28$ simulated SD across the two arms, idiosyncratic spread
$0.41$ and $0.19$; \texttt{share\_significant} mis-calibrated in both
($\mathrm{sd}(z)\approx2.1$), so we use it only on real data \\[2pt]
Contaminated evaluation as a positive control &
Zero-effect null: median $+0.83$ north-star SE, positive for all $69$
candidates, against $+0.0008$ on disjoint folds; $2.26\times$ inflation on
real data \\[2pt]
Pre-declared candidate subset & Instrument association $0.651 \to 0.599$
like for like (Section~\ref{sec:result}). In the held-out comparison
(Section~\ref{sec:horserace}), \texttt{ols\_r}, the statistic the added
candidates were selected on, falls from $0.249$ to $0.191$. The observational
measures' lead over it widens ($+0.056 \to +0.071$), \texttt{pda} and
\texttt{tc\_rho} fall with it and trail by as much as before
($-0.125 \to -0.128$; $-0.123 \to -0.124$), and only
\texttt{deming\_cov\_slope}'s small gap changes sign ($-0.036 \to +0.043$) \\
\bottomrule
\end{tabular}
\end{table}

\paragraph{Where candidate selection matters.} Section~\ref{sec:archive}
disclosed that 20 of the 69 candidates were selected on an uncorrected
observed correlation. This is the one result on which that choice leaves a
visible mark. When the comparison is re-run on the 49 pre-declared candidates
($42.7$ scored per split instead of $58.0$), the uncorrected baseline
\texttt{ols\_r} falls from $0.249$ to $0.191$ and slips from fifth to sixth,
behind both Deming variants (\texttt{deming\_slope} $0.296$,
\texttt{deming\_cov\_slope} $0.234$). That is the direction one expects if
selection modestly flattered the statistic it selected on. The results on
which our conclusion rests either hold or strengthen: the two purely
observational measures stay in the top three
(\texttt{obs\_within\_experiment\_corr} $0.262$ and
\texttt{obs\_within\_experiment\_cov} $0.260$, behind only
\texttt{deming\_slope}), and \texttt{tc\_rho} $(0.067)$ and \texttt{pda}
$(0.063)$ still trail, further behind the observational measures than before. The two most aggressive
corrections do not gain from removing the candidates selected on the
uncorrected correlation, which is the outcome that would have contradicted our
reading.

\paragraph{Transfer across product domains.} Transfer across the archive's two
product domains is mixed (Appendix~\ref{app:domain}). Ranked on the smaller
domain and scored on the larger, \texttt{pda} and \texttt{tc\_rho} again rank
below every uncorrected and observational measure, but \texttt{ols\_r} moves
ahead of both observational measures. The reverse direction, scored on only
$78$ experiments, mostly measures noise. With one deterministic split per
direction, we read this as a caution rather than a finding: a proxy
validated on one product domain should not be assumed to transfer to
another, whether or not it is corrected.

\subsection{Robustness checks}
\label{sec:controls}

Because our conclusion runs against a well-motivated literature, we
pre-specified a set of robustness checks, each designed to fail if the
conclusion were an artifact of implementation, calibration or evaluation
design. Table~\ref{tab:controls} lists them with their results; the
conclusion survives all of them.

\section{Implications for practice}
\label{sec:practice}

Our recommendation is not ``do not correct.'' It is that whether to correct is
an empirical question, which a platform can answer in advance, on the archive
it already has, using three inexpensive diagnostics.

\subsection{Three diagnostics, in order}

\begin{keybox}{Before choosing a proxy-selection estimator}
\small
\textbf{1. Compute your effective archive size.} Not $K$ alone. For each
candidate, compute the effective leverage count of the cross-experiment fit:
the inverse Herfindahl index of the squared deviations of the proxy effects
from their mean (Section~\ref{sec:neff}). If the
median $\neff$ is in the tens, you are in the finite-sample regime of this paper
regardless of how many experiments you have run. If it is also a small
fraction of your $K$, as on ours ($10$ of $262$), your archive's
heterogeneity resembles the one Figure~\ref{fig:crossover} is calibrated to,
and you can read your position off that map from your nominal $K$ and your
north star's reliability. A median $\neff$ much
closer to your $K$ means your archive is more balanced than the one the map is
calibrated to, and for you the map errs on the side of not correcting
(Section~\ref{sec:neff}). \emph{Cost: one pass
over the archive per candidate; no model is fitted.}

\smallskip
\textbf{2. Run the cross-half instrument.} Inside each experiment, split the
customers into two disjoint halves and estimate both the candidate's effect and
the north star's effect on each half; combine them across halves so the two
sides' sampling errors are independent, then measure
the rank association between cross-half agreement and the within-experiment
sampling-error correlation. If it is strongly positive, the sampling-error
covariance is rank-informative on your archive, and with a noisy north star
at effective sizes in the tens, removing it will cost you ranking accuracy. If it is near zero,
the premise of the de-noising literature holds for you, and
Table~\ref{tab:advice} gives the correction to use at your effective size.
\emph{Cost: one extra pass over the archive; no model is fitted.}

\smallskip
\textbf{3. Score candidates on a contamination-free held-out reward.} Rank on
one set of experiments, score on a disjoint set, with non-overlapping customer
folds within experiments \citep{Chou2025}. Report the contaminated version
beside it as a labeled positive control so that the size of the contamination
channel on your platform is visible. \emph{Cost: one additional scoring pass.}
\end{keybox}

Table~\ref{tab:advice} summarizes what to do with the answers. One further
result bears on this advice: the asymptotic ranking accuracy of the Deming
corrections falls, rather than rises, as the north star is measured more
precisely, so a platform that improves the measurement of its north star
should not expect a Deming-based ranking to improve with it
(Appendix~\ref{app:dgp}).

\begin{table}[b]
\centering
\small
\caption{Reading the diagnostics. The recommendation depends on the regime,
and the diagnostics identify it. The effective sizes in the first column
assume a north star measured as noisily as ours; with a better-measured one,
correcting pays at smaller archives (Figure~\ref{fig:crossover}).}
\label{tab:advice}
\begin{tabular}{@{}P{0.22\columnwidth}P{0.22\columnwidth}P{0.44\columnwidth}@{}}
\toprule
\textbf{Effective size} & \textbf{Instrument} & \textbf{Recommendation} \\
\midrule
Tens & Strongly positive &
Rank on an uncorrected or mildly corrected slope. Aggressive de-noising will
cost ranking accuracy. \\[2pt]
Tens & Near zero &
The premise holds, but the sample is too small to support aggressive
correction; prefer the mildest correction and re-test as the archive
grows. \\[2pt]
Hundreds+ & Strongly positive &
Correct, but validate on a ranking loss, not on RMSE; expect the gain to be
small. \\[2pt]
Hundreds+ & Near zero &
Correct. This is the regime for which the de-noising literature was
developed, and the corrections work there. \\
\bottomrule
\end{tabular}
\end{table}

\subsection{What would overturn the advice}
\label{sec:falsify}

Because the advice in Table~\ref{tab:advice} depends on the regime, it would
be overturned not by any single finding about the corrections but by a
mismatch: an archive on which the diagnostics point to one row of the table
while a contamination-free held-out comparison favors the estimator that
another row recommends. Two such mismatches would be especially informative,
because each would refute one of the two claims on which the table rests.

The first is the link between our two legs: that the rank information the
cross-half instrument detects is what makes de-noising costly. Where the
instrument returns a rank association near zero, the uncorrected baseline
should therefore lose its edge over the estimators that remove $\Om$, as it
does when $\Om$ is set to zero in the ablation of
Section~\ref{sec:ablation}. An archive with at least our effective size and
north-star reliability on which the instrument is near zero, yet those
estimators still trail the uncorrected baseline, would sever the link: the
ranking cost would then have another source, and a near-zero instrument would
no longer be a reason to correct.

The second is the crossover map itself. It predicts that the corrections win
on archives with a well-measured north star and an effective size in the
hundreds, even when the shared error is rank-informative, as on ours
(Figure~\ref{fig:crossover}). An archive in that regime on which they still
lose would show that the crossover lies further out than the map places it.
We expect the corrections to win there, and we regard such an archive as the
more useful next study, because it probes the map on the side our own archive
cannot reach. Both tests can be run with the three diagnostics above.

\section{Limitations}
\label{sec:limits}

\paragraph{One archive, one platform.} Every real-data result rests on a single
archive from a single consumer platform with a single north star. The
simulation generalizes the mechanism across reliability and archive size, but
its calibration comes from this archive. The central empirical claim, that
$\Om$ is rank-informative, is a claim about this archive, and it might not hold
on every platform: a recent industry workshop on long-term evaluation found
many of its conclusions to depend on context, such as the type of platform
\citep{Sigerson2026}. No platform has to take it on trust, however. The
cross-half instrument (Section~\ref{sec:instrument}) tests it directly on any
archive with customer-level data, and the paper's contribution is as much that
test as the answer it gives here.

\paragraph{No economic magnitude.} A difference can matter at three levels:
statistical detectability, ranking consequence and economic importance. We
establish the first two but not the third. Converting a 17-point pair-accuracy
gap, or a $3.4$-fold top-5 regret gap, into revenue would require the value of
the decisions taken downstream of the shortlist.

\paragraph{The simulation is a model.} The generator reproduces the share of
experiments in which a candidate reaches significance only on average, not
candidate by candidate (Section~\ref{sec:calibration}), so we use that
statistic only as measured on the real archive. It also understates dependence
between candidates: within a replicate, candidates
share the experiments and the north-star shock but draw their proxy-side noise
independently, whereas 55 of our 69 are equal-weight combinations of the same
14 base metrics. Whole-catalog pair accuracy, a sum over pairs, is fairly
robust to this; head-of-list losses, which depend on the joint distribution of
near-ties, are not. This is a limit of the model, not of the calibration,
which targets marginal moments by construction (Table~\ref{tab:calib}), and it
is why we regard our head-of-list results as the least settled in the paper,
although they replicate across both calibration arms.

\paragraph{The held-out reward measures gain per shipped decision.} It averages
the north-star effect over firing events, the cases in which the rule ships, so
it targets $\mathrm{E}[\tau^{y} \mid \text{fired}]$ rather than the
unconditional policy value $\mathrm{E}[\tau^{y} \cdot \mathbf{1}\{\text{fire}\}]$,
the criterion of \citet{Chou2025}. The two can rank rules differently: a rule that
fires rarely but gains a lot each time wins on the conditional criterion and
can lose on the unconditional one. Because every measure is scored against the
same reward, the comparison between measures is largely insulated from this
choice, though not completely, since firing rates differ across candidates.
The reward is also a ratio, the summed north-star effect divided by the number
of firing events, and like any ratio estimator it is consistent but biased in
finite samples.

\paragraph{The 69 candidates are not 69 independent tests.} They share a small
number of components, so candidate-resampled intervals are optimistic, by an
amount we do not correct for. On the 14 base metrics alone, the instrument
keeps its sign and roughly three-quarters of its size but is no longer
distinguishable from zero (Section~\ref{sec:result}), which is to be expected
with so small a sample. The experiment-resampled evidence of
Section~\ref{sec:horserace} does not share this limitation.

\paragraph{Two circularities inside the simulation.} The generator's truth is an
estimator's own output (\texttt{re\_rho} in one calibration arm,
\texttt{tc\_rho} in the other), which favors those estimators in the arm that
uses them; reporting both arms mitigates this, and any residual advantage goes
to the corrections. The second circularity runs the other way: the simulated
\texttt{obs\_corr} is close to a re-expression of the calibration's own
$\rho$--$\Om$ association and is invariant to the reliability lever by
construction (Appendix~\ref{app:dgp}), so its simulated accuracy is not
independent evidence for our claim, and its flatness across rungs is a
property of the construction, not a finding. The independent evidence comes
from the real data: the cross-half instrument of Section~\ref{sec:instrument}
and the held-out reward of \texttt{obs\_corr}'s real-archive counterpart,
\texttt{obs\_within\_experiment\_corr}, in Section~\ref{sec:horserace}.

\paragraph{Estimator coverage.} The estimators we study span the dose gradient
but not every proposal in a fast-moving literature; in particular, they do not
include the full hierarchical model of \citet{Tripuraneni2024} or the Bayesian
multilevel treatment of \citet{Gazvoda2024} in its native form. Our claims are
stated in terms of the dose of $\Om$ removed, which we regard as the operative
variable.

\section{Conclusion}
\label{sec:conclusion}

Many product decisions on online platforms rest on proxy metrics, which makes
choosing those proxies well one of the central measurement problems in online
experimentation. Recent work has established that the within-experiment
covariance of sampling error biases estimates of the cross-experiment
covariance of true treatment effects, a common basis for that choice, and has
developed effective estimators that remove it. We asked a different question:
does removing it improve the ranking of proxy candidates that a platform acts
on?

The answer depends chiefly on three properties of the archive, each of which a
platform can measure on its own data: whether the shared error carries rank
information, how reliably the north star is measured and how many experiments
effectively inform the ranking. On our archive of 262 randomized experiments,
all three point against removing it. Across candidates, the quantity being
removed has a Spearman correlation of $0.65$ with a model-free measure of true
alignment that shares no sampling error, cross-experiment fit or subtraction
with it: on this archive, it is rank-informative. With a noisy north star and
a small effective archive size, removing it therefore costs ranking accuracy,
and in a simulation calibrated to the archive's own moments the cost scales
with the dose removed, from the mildest correction, which outperforms the
uncorrected baseline, to the aggressive de-noise-and-rescale estimator, which
trails it by seventeen points. The cost survives exact subtraction, so better estimates of
$\Om$ cannot remove it, and a counterfactual ablation confirms the mechanism:
once $\Om$ is set to zero, the baseline falls behind all but one of the
corrections. The corrections behave as designed, but they assume that the
shared error carries no information about proxy quality, and on this archive
it does.

The practical consequence is a map. Correcting pays above a crossover in
archive size that rises steeply as north-star reliability falls, and north
stars, chosen because they measure something slow and consequential, tend to
be among the least reliably measured metrics a platform has. At every
north-star reliability rung we simulate at or below $0.60$, with our archive's
structure held fixed, the two most aggressive corrections overtake the
uncorrected baseline only once on our grid, by less than half a point at
$K = 1000$, and the published archives whose size we could verify, with $123$
to $307$ experiments, lie well below that size. Measured by leverage,
our own archive has an effective size of about ten experiments for the median
candidate. On held-out experiments, a contamination-free comparison reproduces
the sign of six of the seven predicted gaps; the seventh is a purely
observational measure that ranks higher on the real archive than the
simulation predicted.

We therefore offer three diagnostics rather than a verdict: measure your
effective archive size, run the cross-half instrument to find out whether the
premise holds on your data, and evaluate estimators on a contamination-free
held-out ranking loss, not on estimation error. Each needs little more than
one pass over data a platform already holds. The natural next step, which we
leave for future research, is to run them on archives with well-measured north
stars and effective sizes in the hundreds, where the map predicts that the
corrections should win; finding that they do would mark the boundary of our
result from the other side.

The broader lesson is that an estimator can be right for the question it was
built to answer and still be the wrong tool for the decision that follows from
it. For proxy selection, that lesson takes a concrete form:
on archives like this one, correlated sampling error is not
contamination to be removed but signal to be ranked on, and whether to discard
it should be decided by measurement rather than by assumption.


\appendix

\section{Estimators}
\label{app:estimators}

Every estimator takes the same inputs for one candidate: the per-experiment
effect estimates $(\hat\tau^{x}_k,\hat\tau^{y}_k)$, $k = 1,\dots,K$, and their
sampling covariance matrices $\Omega_k$. Each returns one score, and
candidates are ranked by it. The real archive and its bootstrap resamples are
scored by the same code, and the simulations by a copy of it that is checked
against the real-archive estimates before anything is simulated.

\paragraph{Notation.} Let $s_{xx}$, $s_{yy}$ and $s_{xy}$ be the sample
variances and covariance of the effect estimates across the $K$ experiments,
with divisor $K-1$, and let $\bar\omega^{xx}$, $\bar\omega^{yy}$ and
$\bar\omega^{xy} = \Om$ be the unweighted archive averages of the sampling
variances and covariance. The de-noised moments
\begin{equation}
\hat\sigma_x^{2} = s_{xx} - \bar\omega^{xx}, \qquad
\hat\sigma_y^{2} = s_{yy} - \bar\omega^{yy}, \qquad
\widehat{\Sig} = s_{xy} - \Om
\label{eq:denoised}
\end{equation}
are the method-of-moments estimates implied by Equation~\eqref{eq:decomp} and
its analogues for the two variances.

\paragraph{Uncorrected baselines.} \texttt{ols\_r} and \texttt{ols\_slope}
are the correlation $s_{xy}/(s_{xx}s_{yy})^{1/2}$ and the slope
$s_{xy}/s_{xx}$ of an ordinary least-squares fit of $\hat\tau^{y}$ on
$\hat\tau^{x}$. Both keep $\Om$ in full.

\paragraph{Mild corrections.} \texttt{deming\_slope} is the slope of a Deming
regression \citep{Deming1943} with known, experiment-specific error variances:
the $b$ that, together with an intercept $a$, minimizes
\begin{equation}
\sum_{k=1}^{K}\frac{(\hat\tau^{y}_k - a - b\,\hat\tau^{x}_k)^{2}}
{\omega^{yy}_k + b^{2}\omega^{xx}_k}.
\label{eq:deming}
\end{equation}
The $b^{2}\omega^{xx}_k$ term corrects the attenuation that the proxy's
sampling variance causes in a least-squares slope; $\omega^{xy}_k$ does not
enter, so $\Om$ is kept. \texttt{deming\_cov\_slope} is York's extension to
correlated errors \citep{York1968}, which replaces the denominator of
Equation~\eqref{eq:deming} by $\omega^{yy}_k - 2b\,\omega^{xy}_k +
b^{2}\omega^{xx}_k$, the sampling variance of $e^{y}_k - b\,e^{x}_k$ when the
two errors covary. It takes $\Om$ into account rather than subtracting it.
Both fits assume that the true effects lie exactly on a line. When they do
not, as whenever $|\rho| < 1$, York's slope still depends on $\Om$ in the
limit, which is why we describe it as removing $\Om$ only partially. The
surrogate-endpoint model of \citet{DanielsHughes1997} relaxes this assumption by
adding a between-trial residual variance; its slope then targets
$\Sig/\sigma_x^{2}$, as \texttt{tc\_slope} does. In both of our
fits the intercept is profiled out as a weighted mean, and the slope is found
by searching over the angle of the line rather than the slope itself, which
keeps the search range bounded however steep the line: a grid of $361$ angles
locates the minimum, and a bounded one-dimensional search refines it within
one grid step. On the real archive both fits agree with the reference
implementations in the R packages \texttt{deming} \citep{Therneau2024} and
\texttt{IsoplotR} \citep{Vermeesch2018} to within $3\times10^{-5}$ and
$2\times10^{-6}$ respectively, with identical candidate rankings.

\paragraph{Full de-noising by moments.}
$\texttt{tc\_slope} = \widehat{\Sig}/\hat\sigma_x^{2}$ is the total-covariance
slope of \citet{Bibaut2024}, and
$\texttt{tc\_rho} = \widehat{\Sig}/(\hat\sigma_x\hat\sigma_y)$ the correlation
read off the same de-noised moments. \texttt{tc\_slope} is
undefined when $\hat\sigma_x^{2} \le 0$. \texttt{tc\_rho} is undefined when
either de-noised variance is non-positive, and inadmissible when the ratio
falls outside $[-1, 1]$; both cases are treated as missing
(Section~\ref{sec:tcrho}). Because $\hat\sigma_y^{2}$ is computed from the
north star alone, it is shared by all candidates up to the occasional missing
experiment, so when it is non-positive \texttt{tc\_rho} fails for essentially
every candidate at once.

We form the cross-experiment moments with the unbiased $K-1$ divisor. Run on
the authors' own simulation design, the resulting estimator reproduces the
published code of \citet{Bibaut2024} to within $1.7\times10^{-12}$ on the
estimated slope, and at $n = 5{,}000$ customers per arm it reduces the naive
slope's bias from $+0.341$ to $-0.017$, as designed.

Our own construction, $\texttt{pda} = \widehat{\Sig}/\hat\sigma_x$, is the
de-noised slope rescaled to a one-standard-deviation change in the
candidate's true effect: it equals $\texttt{tc\_slope}\cdot\hat\sigma_x$, or
equivalently $\texttt{tc\_rho}\cdot\hat\sigma_y$. It removes $\Om$ in full;
the ``partially'' in its name, \emph{partially de-noised alignment}, refers to
the normalization: only the candidate's spread is de-noised and divided out.
Its limit is $\rho\,\sigma_y$. Because every candidate is held against the
same north star, $\sigma_y$ is common to all of them, so \texttt{pda}, like
\texttt{tc\_rho}, is rank-faithful in the limit: it orders candidates exactly
by $\rho$. \texttt{tc\_slope}, whose limit $\rho\,\sigma_y/\sigma_x$ also
depends on the candidate's own spread, is not. As \texttt{pda} never uses
$\hat\sigma_y$, it is undefined only when $\hat\sigma_x^{2} \le 0$. It is not
a correlation and is not bounded by one.

\paragraph{Full de-noising by likelihood.} \texttt{re\_rho} is the
true-effect correlation of a bivariate random-effects meta-analysis
\citep{vanHouwelingen2002}, which models each experiment's pair of estimates as
\[
(\hat\tau^{x}_k,\hat\tau^{y}_k)^{\top} \sim \mathcal{N}(\mu,\ G + \Omega_k),
\]
with $\mu$ the two mean effects and $G$ an unstructured $2\times2$ covariance
matrix of the true effects. Because the known $\Omega_k$ enter the likelihood
directly, $\Om$ is modeled away rather than subtracted. We estimate $G$ by
restricted maximum likelihood and report
$\texttt{re\_rho} = \hat G_{12}/(\hat G_{11}\hat G_{22})^{1/2}$. The search runs over the
Cholesky factor of $G$ with log-transformed diagonal, so that every value it
visits is a valid covariance matrix, and uses Nelder--Mead from two starting
points with zero correlation, built from the de-noised variances (floored at
$10^{-10}$) and from the raw variances of the estimates. The better fit is
kept and the estimate is missing if both fail. Bootstrap refits start from
the full-sample solution. This is the model that \texttt{rma.mv} in the R
package \texttt{metafor} \citep{Viechtbauer2010} fits with an unstructured
random-effects covariance (\texttt{struct="UN"}, \texttt{method="REML"}).
On the real archive the two agree to within $3\times10^{-5}$ in
\texttt{re\_rho} and $10^{-7}$ in the maximized restricted log-likelihood,
with identical candidate rankings.
\citet{Tripuraneni2024} fit a Bayesian version of this model to the north star
and all proxies jointly, with multivariate-normal true effects, and
\citet{Gazvoda2024} propose a version with bivariate Student-$t$ true effects;
neither is part of our comparison (Section~\ref{sec:limits}).

\paragraph{Reference statistics.}
$\texttt{obs\_corr} = \Om/(\bar\omega^{xx}\bar\omega^{yy})^{1/2}$ uses only
the sampling covariance matrices. In the held-out comparison of
Section~\ref{sec:horserace}, \texttt{obs\_within\_experiment\_corr} is the same
statistic computed on the ranking half, and
\texttt{obs\_within\_experiment\_cov} is its numerator alone.
\texttt{share\_significant} is the share of experiments in which the
candidate's effect is significant at the two-sided 5\% level, computed on the
ranking half in the held-out comparison. It measures sensitivity rather than
alignment, and it is the one statistic our calibration audit finds clearly
mis-calibrated (Section~\ref{sec:calibration}).

\paragraph{$t$-statistics and oracle benchmarks.} The primary simulation also
scores eight measures on which none of our results rests.
\texttt{ols\_tstat}, \texttt{deming\_tstat}, \texttt{deming\_cov\_tstat} and
\texttt{tc\_tstat} divide the corresponding slope by its standard error from an
inner bootstrap of $400$ resamples of the replicate's experiments;
\texttt{ols\_tstat\_homoskedastic} uses the closed-form least-squares standard
error instead. The three \texttt{\_oracle} variants of the Deming, York and
total-covariance $t$-statistics divide the slope by its standard deviation
across the $300$ independent replicates of the simulation, that is, by its true
sampling variability. No analyst can compute this denominator: it would
require many independent archives drawn from the same population, whereas a
platform observes only one.
They are diagnostic benchmarks: they isolate how much of a $t$-statistic's
performance is lost because it must estimate its own precision, as
distinct from the estimator itself.

\paragraph{Coverage by analysis.} The primary simulation scores fifteen
measures: the seven estimators of Table~\ref{tab:dose} other than
\texttt{ols\_slope}, and the eight above. The reliability sweep, which also
supplies the ablation of Section~\ref{sec:ablation}, scores seven:
\texttt{ols\_r}, \texttt{deming\_slope}, \texttt{deming\_cov\_slope},
\texttt{tc\_slope}, \texttt{tc\_rho}, \texttt{pda} and \texttt{obs\_corr}. It
omits \texttt{re\_rho} for computational reasons: the random-effects fit
requires a numerical likelihood maximization for every candidate in every
simulated archive, and the sweep simulates more than ten times as many
archives as the primary simulation. Figure~\ref{fig:simreal} therefore takes
\texttt{re\_rho} from the primary simulation at the same archive size and
calibration arm. The two grids also draw their archives differently: the sweep
draws experiments with replacement, whereas the primary simulation draws them
without replacement and so uses exactly the archive's experiments at
$K = 262$. Their accuracies for the same estimator differ slightly,
by at most $0.7$ points at $K = 262$ (Tables~\ref{tab:dose2}
and~\ref{tab:losses}). The held-out comparison on the real archive scores eleven
measures: the eight estimators and the three ranking-half statistics above.

\section{Simulation specification}
\label{app:dgp}

Each replicate draws $K$ experiments. True effects
$(\tau^{x}_k,\tau^{y}_k)$ are drawn from a distribution calibrated to the
archive's cross-experiment effect covariance. Because the archive's effects
are heavy-tailed, a normal distribution with that covariance would make too
many experiments significant, so we use a scale mixture of bivariate normals:
one experiment in twenty comes from a wider component, whose width is a single
number, common to all candidates, set so that the share of significant
experiments averaged over candidates matches the archive's. Whether an
experiment is wide is drawn once and applies to the north star and every
candidate alike, and the mixture is normalized to the calibrated covariance,
so the true correlations stay exactly at their calibrated values. Sampling
errors are drawn as $\mathcal{N}(0,\Omega_k)$, with $\Omega_k$ the measured
sampling covariance matrix of the drawn experiment, so that the joint
distribution of precision across experiments is preserved. Every
$\Omega_k$ is checked for positive semi-definiteness before use, and the
simulation stops if any check fails.

The generator runs on two grids. The primary simulation draws experiments
without replacement, at $K \in \{20,50,100,200,262\}$ and in both calibration
arms, so that at $K = 262$ it uses exactly the archive's own experiments; its
north star keeps the reliability that the generator implies at the measured
moments. The reliability sweep draws experiments with replacement, which lets
it reach $K = 1000$, and crosses five archive sizes
$K \in \{20,50,100,262,1000\}$, five
north-star reliability levels $\{0.10,\,0.104,\,0.373,\,0.60,\,0.951\}$ (the
second being the archive's measured value), two calibration arms
(random-effects and tail-correlation) and two sampling-covariance conditions
(as measured and set to zero). North-star reliability is varied by scaling the
north-star sampling variance by a factor $f$ and its sampling covariance with
each candidate by $\sqrt{f}$, which leaves every sampling-error correlation,
and hence \texttt{obs\_corr}, unchanged. The true-effect distribution is held
fixed, so that reliability is manipulated without changing the estimand.

The rungs are fixed before the generator is calibrated, so the rung for our
archive, $0.1038$, is not identical to the generator's own implied reliability
at the measured moments, $0.1249$. Running the archive at the rung rather than
at its own value inflates the north star's sampling variance by a factor of
$1.23$. The change is small, but it is not conservative: it places our archive
marginally further below the crossover than the generator's own calibration
would. We quote $0.104$ for the rung throughout and $0.125$ where we report
the random-effects estimate itself.

Asymptotic ceilings are computed analytically from the same DGP parameters by
evaluating each estimator's probability limit per candidate and scoring the
resulting ranking. Because ceilings and finite-sample accuracies are computed
on the same grid, they are directly comparable, and we do not compare either
with numbers from a different simulation. The two Deming ceilings are
approximations rather than identities: their probability limits are evaluated
at the mean per-experiment sampling variance across the archive, which
gives the exact limit only under homoskedastic sampling error. Our archive is
strongly heteroskedastic (Section~\ref{sec:neff}), and we cannot sign the
direction of the resulting error. This is the most likely explanation
for the finite-sample cells that sit above their own ceiling, all of which
are Deming cells. At $K = 262$ there are five, above by $0.04$ to $0.20$
(Section~\ref{sec:inversion}), all at reliability $0.951$, where the
approximate ceilings are near or below chance. The same five cells exceed
their ceilings at every other size of the sweep, and at $K = 1000$ five more do
so by at most $0.018$, all in the tail-correlation arm at reliabilities
$0.373$ and $0.60$. The ceilings of the uncorrected baseline and of the
total-covariance and random-effects estimators are exact under the DGP and are
never exceeded, at any size.

Two by-products of the reliability sweep bear on the practical advice of
Section~\ref{sec:practice}, and we state them separately because they are easy
to get backwards.

\paragraph{Deming is not rank-neutral in north-star reliability.} North-star
attenuation is exactly rank-neutral for \texttt{tc\_slope} and \texttt{pda},
and for the uncorrected baseline when the sampling covariance is zero. It is
not neutral for the Deming variants, whose slope-dependent weights
carry the north-star sampling variance into the objective, and the effect runs
the wrong way. \texttt{deming\_cov\_slope}'s asymptotic
ranking gets monotonically worse as the north star is measured better,
from $0.930$ at reliability $0.10$ through $0.882$ at $0.373$ and $0.845$ at
$0.60$ to $0.542$ at $0.951$: gradually over the range real archives occupy,
steeply at the top. A platform that measures its north star more precisely
should therefore not expect its Deming-based ranking to improve. As an
internal check, with $\Om$ ablated to zero the two Deming variants become the
same estimator, and their ceilings coincide at every rung, as they should.

\paragraph{The covariance channel changes sign with north-star reliability.}
The uncorrected baseline ranks on a mixture of a true-effect signal and
a sampling-covariance signal, and whether the second helps its asymptotic
ranking depends on how well the north star is measured. Switching the sampling
covariance off fixes its ceiling at exactly $0.9126$ at every rung
(random-effects arm). Leaving it in gives $0.909$ at our archive's $0.104$, then
$0.922$, $0.925$ and $0.933$ at $0.373$, $0.60$ and $0.951$: below the
covariance-free ceiling at our archive's reliability, above it at every higher
rung, with the crossing between $0.104$ and $0.373$. The tail-correlation arm
shows the same channel with the crossing shifted to between $0.60$ and $0.951$
(its covariance-free ceiling is $0.9355$, which the mixture exceeds only at
$0.951$), so we state the location of the crossing as a random-effects-arm
result and do not generalize it. The practical
reading is the same in both arms, and our argument depends on it: at our
archive's reliability the correction is right asymptotically, which is why the
finite-sample result in Section~\ref{sec:inversion} matters.

\section{Held-out reward protocol}
\label{app:holdout}

Splits partition experiments into a ranking set and a disjoint scoring set.
Each candidate receives two numbers per split. Its \emph{score} is the
alignment measure under test, fitted on the ranking set only. Its
\emph{reward} is the average north-star effect per shipped decision on the
scoring set, under the rule ``ship when the proxy moves significantly and
upwards'': within each scoring experiment customers are partitioned into two
folds, the trigger is read on one fold and the payoff on the other, and the
two directions are pooled over firing events. No payoff is therefore selected
on its own sampling noise. The fold split is the device of
\citet{Chou2025}, applied here to a conditional per-decision payoff rather
than to their cumulative rule return (Section~\ref{sec:limits}). The reported
statistic is the Spearman correlation across candidates between score
and reward, averaged over the scored splits of each cell ($67$ of $100$ in the
headline cell). Percentiles across splits show how much it moves when the
experiments are divided differently; a bootstrap over candidates within each
split, pooled across splits, shows how much it moves when a different set of
candidates is scored. The same-sample
version (trigger and payoff read on the same customers) is retained unchanged
as a labeled positive control.

\paragraph{Ordering within the leading group.} The three-group separation of
Section~\ref{sec:horserace} holds at every random-split ranking size, but the
order within the leading group varies. At ranking size $100$ it is
\texttt{deming\_slope} $0.270$, \texttt{obs\_within\_experiment\_cov} $0.268$,
\texttt{obs\_within\_experiment\_corr} $0.264$ and \texttt{ols\_slope}
$0.260$, with \texttt{ols\_r} at $0.225$.

\paragraph{Convex blend.} For each observational measure, we also fit the
optimal convex blend of that measure and the uncorrected baseline
\texttt{ols\_r} on the evaluation data itself, separately for each splitting
scheme and ranking size. The blend places all its weight on the
observational component in every random-split comparison, and at least $0.8$
of it in the temporal test and when candidates are ranked on the larger product
domain. Only when they are ranked on the smaller domain and scored on the
larger, where \texttt{ols\_r} also ranks ahead of both observational measures
(Appendix~\ref{app:domain}), does most of the weight go to \texttt{ols\_r}. The probe does not handicap the cross-experiment side: \texttt{ols\_r} is the
best-performing cross-experiment measure on the correlation scale, not a
de-noised one. Because the weight is fitted in sample, the probe shows which
component the held-out rewards favor; it is not a recommendation to use such a
blend.

\section{Instrument design sweep}
\label{app:designs}

Table~\ref{tab:designs} re-computes the headline instrument (the rank
association between cross-half agreement and the within-experiment
sampling-error correlation) on subsets of the archive, and under four
values of the minimum number of experiments the pipeline requires before it
computes a candidate's cross-half agreement.
All figures are on the all-candidate basis ($69$ candidates, $68$ where one
candidate is undefined), so the whole-archive row is the $0.686$ of
Section~\ref{sec:result} rather than the $0.651$ computed on the $63$
candidates that every truth definition scores.

The association is positive and substantial in every design. It does not
vary with that threshold ($10$, $20$, $30$ or $40$): every candidate is
measured in every experiment and the smallest subset has $78$ of them, so
even the highest threshold never binds; we report the sweep only to document
this. The
two temporal halves bracket the whole-archive value closely. The two product
domains differ more, consistent with the domain heterogeneity in
Appendix~\ref{app:domain}: the association is strongest on the smaller
domain and weakest on the larger one, and the gap between them is
larger than anything the temporal split produces.

\begin{table}[ht]
\centering
\small
\caption{\datasource{Real archive} The instrument under alternative designs.
$\rho$ is the Spearman rank
correlation with the within-experiment sampling-error correlation. All four
minimum-experiment thresholds ($10$--$40$) give identical values, so a single
column is shown.}
\label{tab:designs}
\begin{tabular}{@{}lccc@{}}
\toprule
\textbf{Design} & \textbf{Exps.} & \textbf{Cands.} & $\boldsymbol{\rho}$ \\
\midrule
Whole archive                & $262$ & $69$ & $0.686$ \\
First half by end date       & $132$ & $69$ & $0.709$ \\
Second half by end date      & $130$ & $69$ & $0.657$ \\
Larger domain only           & $184$ & $69$ & $0.597$ \\
Smaller domain only          & $78$  & $68$ & $0.907$ \\
\bottomrule
\end{tabular}
\end{table}

\section{Influential experiments}
\label{app:influence}

Two experiments dominate the cross-experiment fits of Section~\ref{sec:neff}.
We call them A and B. Both come from the larger domain and from the second half of
the archive by end date; A has $464{,}717$ customers and B $5{,}428{,}826$, and
they are the single highest-leverage experiment for $47$ and $14$ of the $69$
candidates. Table~\ref{tab:influence} repeats the analysis without A, without
B, without both and without each of the $262$ experiments in turn. Every
deletion rebuilds both sides of each association: $\Om$, re-averaged over the
remaining experiments, and true alignment under the four definitions of
Table~\ref{tab:truth} other than cross-half agreement, refitted on the same
experiments. The $\Om$ side barely moves: under every deletion, the ordering
of the candidates by $\Om$ agrees with the full-archive ordering at a Spearman
correlation of at least $0.998$. Cross-half agreement is not refitted, because
it is built from customer-level half samples that this analysis does not read;
Table~\ref{tab:designs} tests it instead on parts of the archive that exclude
both experiments.

\begin{table}[t]
\centering
\small
\setlength{\tabcolsep}{3.5pt}
\caption{\datasource{Real archive} Deleting the experiments that dominate the
cross-experiment fits. A and B are the single highest-leverage experiment for
$47$ and $14$ of the $69$ candidates; \emph{Single} is the range over all
$262$ single-experiment deletions, A and B included. Associations are Spearman
correlations between $\Om$ and true alignment on the $63$ candidates of
Table~\ref{tab:truth}, with both sides refitted after each deletion; the
total-covariance row uses only the candidates on which that estimator remains
admissible. Reliabilities and $\neff$ use all $69$ candidates. RE: random
effects. Our cross-half instrument needs customer-level
data and is tested instead in Table~\ref{tab:designs}.}
\label{tab:influence}
\begin{tabular}{@{}lccccc@{}}
\toprule
 & \textbf{Full} & $-$\textbf{A} & $-$\textbf{B} & $-$\textbf{A,B} & \textbf{Single} \\
\midrule
\multicolumn{6}{@{}l}{\emph{Association with true alignment}} \\
RE, covariance set to zero & $0.760$ & $0.769$ & $0.912$ & $0.910$ & $0.728$--$0.912$ \\
RE, full                   & $0.697$ & $0.707$ & $0.867$ & $0.862$ & $0.607$--$0.867$ \\
Total covariance           & $0.666$ & $0.681$ & $0.764$ & $0.769$ & $0.108$--$0.764$ \\
\quad admissible           & $63$    & $63$    & $63$    & $63$    & $9$--$63$ \\
Uncorrected \texttt{ols\_r} & $0.886$ & $0.883$ & $0.933$ & $0.932$ & $0.863$--$0.933$ \\
\addlinespace
\multicolumn{6}{@{}l}{\emph{Reliability}} \\
North star, moments        & $0.104$ & $0.110$ & $0.105$ & $0.111$ & $0.014$--$0.130$ \\
North star, RE             & $0.125$ & $0.124$ & $0.128$ & $0.126$ & $0.071$--$0.130$ \\
Candidates, median         & $0.759$ & $0.690$ & $0.755$ & $0.679$ & $0.690$--$0.768$ \\
Candidates, minimum        & $0.331$ & $0.287$ & $0.332$ & $0.288$ & $0.270$--$0.358$ \\
\addlinespace
Median $\neff$             & $10.0$  & $17.8$  & $13.9$  & $22.6$  & $9.2$--$17.8$ \\
\bottomrule
\end{tabular}
\end{table}

\paragraph{The two dominant experiments.} Removing A changes each association
by less than $0.02$. Removing B, or both, raises all four by between $0.05$
and $0.17$, to between $0.77$ and $0.93$ without both. Far from carrying the
association between $\Om$ and true alignment, then, the two experiments that
dominate the fits weaken it, and B accounts for nearly all of that.

\paragraph{Single deletions.} The association is robust to every single
deletion under both random-effects definitions and for the uncorrected
baseline: none of the $262$ deletions lowers it by more than $0.10$. The
total-covariance definition is the only exception, and it fails as
Section~\ref{sec:tcrho} would predict. Two other experiments carry much of the
north star's signal, and deleting either one leaves so little of it that
\texttt{tc\_rho} falls outside $[-1, 1]$ for half the candidates or more and
its association with $\Om$, computed on the candidates that remain,
collapses. The failure is the estimator's, not the association's: on the same
reduced archives, and even on the same remaining candidates, the other three
definitions keep a clear positive association with $\Om$. Every other
deletion leaves the total-covariance association at $0.545$ or above.

\paragraph{Calibration.} The reliability that places the archive on the
crossover map is the north star's, and no deletion raises it by more than
$0.03$: both estimates stay at or below $0.130$, against $0.104$ and $0.125$
on the full archive. Deletions that lower it only move the archive further
from the crossover. The north star also remains the least reliable of the $70$ metrics.
The least reliable candidate never falls below $0.270$, and the median
candidate remains more than five times as reliable as the north star under
either estimate. The candidates' reliabilities, which the simulation also
inherits, move more: removing A lowers their median from $0.759$ to $0.690$,
because A carries the most extreme proxy effect for $47$ candidates. We do not
rerun the simulation at these values. The held-out comparison, discussed
below, tests its prediction at ranking sizes at which most ranking sets
exclude both experiments.

\paragraph{Effective size.} The caveat of Section~\ref{sec:neff} applies to
the archive without A and B. Its median $\neff$ is $22.6$, above the $13.1$ of
a simulated archive of nominal size $262$ but below the $40.7$ of one of size
$1000$. For this archive, reading Figure~\ref{fig:crossover} at nominal size
errs toward not correcting; read by effective size, it lies between those two
sizes. At the reliability rung we use for our archive, $0.104$,
\texttt{tc\_slope} overtakes the uncorrected baseline at neither, and the
reduced archive's own reliability estimates, $0.111$ and $0.126$, lie within
$0.01$ of the full archive's.

\paragraph{Held-out comparison.} The held-out rewards need customer-level
folds and are not recomputed here either, but the design limits how much A
and B can affect them. The concentration that $\neff$ measures is a property
of the cross-experiment fits, which the held-out comparison makes on the
ranking set only; on the scoring set, each experiment contributes at most one
decision per candidate. A random ranking set of $s$ experiments contains
neither A nor B with probability $(262-s)(261-s)/(262\cdot 261)$, which is
$0.853$, $0.654$, $0.381$ and $0.249$ at the four ranking sizes of $20$, $50$,
$100$ and $131$. The separation between the three groups of measures in
Section~\ref{sec:horserace} holds at every size: at size $20$, where most
ranking sets exclude both experiments, as at size $131$, where three in four
include at least one.

\section{Candidate selection and the Berkson check}
\label{app:selection}

Section~\ref{sec:archive} explains why we added 20 candidates to the 49
pre-declared ones and why their selection could bias the instrument in either
direction. This appendix describes the selection rule, reports how stable its
picks are and compares the instrument between the two groups.

\paragraph{The selection rule.} The rule builds candidates from $12$ of the
$14$ individual metrics, leaving out one redundant ratio and the profit
metric. The profit metric covers only the experiment window and is therefore
itself a component of the north star, so any combination containing it would
gain a mechanical head start on directionality. The rule then proceeds in
three steps. It forms all $2{,}497$ equal-weight combinations of two to six of
the $12$ metrics. It draws $600$ bootstrap resamples of the experiments and
counts how often each combination lies on the Pareto front of sensitivity
against directionality. Finally, it ranks the combinations by that count and,
going down the ranking, keeps each one whose components do not overlap too
heavily with those of a candidate already in the catalog (Jaccard similarity
at most $0.70$), until $20$ are kept. The profit metric itself stays in the
catalog as one of the $14$ individual metrics; dropping it moves the
instrument only from $0.686$ to $0.688$.

\paragraph{Stability of the picks.} To measure how much the $20$ picks owe to
this particular sample, we re-ran the rule on $200$ random halves of the
experiments and recorded how often each of the $20$ was picked again. Six
return in more than half of the splits, the most stable in $79\%$ of them,
but the median pick returns in only $30\%$: far more often than the one split
in a hundred or so that a blind pick of $20$ from these combinations would
manage, yet far from reliably. With an effective archive size in the tens
(Section~\ref{sec:neff}), the data are too thin for any rule of this kind to
settle on the same $20$ of $2{,}497$ combinations. The particular $20$ are
therefore partly an accident of this sample, one more reason to rest the
argument on the pre-declared candidates.

\paragraph{The Berkson check.} The rule favored candidates with high
directionality, the uncorrected observed correlation with the north star, and
that correlation rises with both quantities the instrument relates: the
alignment of the candidate's true effects with the north star's and the
correlation of their sampling errors. Selecting on a quantity that rises with
both makes the two negatively dependent among the selected candidates, because
a candidate that is low on one must be high on the other to be picked. This is
Berkson's paradox \citep{Berkson1946}, and it predicts that the instrument
should be weaker among the $20$ selected candidates than among the $49$
pre-declared.

On the full catalog the data show this pattern, though only weakly: the
instrument is $0.571$ among the selected candidates and $0.626$ among the
pre-declared. Both lie below the $0.686$ on all $69$ candidates. This reflects
the other effect of selection noted in Section~\ref{sec:archive}: it placed
the selected candidates high on both quantities, so the two groups are offset
along the direction of the association, and pooling them adds between-group
spread that is aligned with it. On the $63$ candidates that every truth
definition scores, the within-group comparison reverses, with $0.738$ among
$16$ selected and $0.599$ among $47$ pre-declared candidates, but with only
$16$ candidates the bootstrap interval for the selected group runs from $0.28$
to $0.92$. On the covariance scale of Section~\ref{sec:result}, the
association among the $20$ selected candidates is flat ($-0.075$,
$p = 0.75$). Subgroups this small cannot settle the direction of the
within-group effect, which is why Section~\ref{sec:archive} relies instead on
reporting every headline result on the $49$ pre-declared candidates alone.

\section{Calibration audit}
\label{app:calib}

\begin{table}[t]
\centering
\small
\caption{\datasource{Simulation vs.\ real archive} Calibration audit.
\emph{Cov.}: share of candidates whose real-archive statistic falls inside the
simulated 90\% interval (nominal $90\%$). $z$: standardized departure of the
observed statistic from the simulated median,
$(\text{observed} - \text{median}) / \hat\sigma$ with
$\hat\sigma = (\text{hi} - \text{lo})/(2 \times 1.645)$; $\bar z$ and
$\mathrm{sd}(z)$ are its mean and standard deviation across candidates.
$\dagger$: on the $63$ candidates where \texttt{tc\_rho} is defined; $0.81$
if the six undefined cases count as misses. $\ddagger$: close to
tautological, because this arm's generator takes its true correlations from
this estimator; the other arm gives the informative reading.}
\label{tab:calib}
\begin{tabular}{@{}lccc@{}}
\toprule
\textbf{Statistic} & \textbf{Cov.} & $\bar z$ & $\mathrm{sd}(z)$ \\
\midrule
\multicolumn{4}{@{}l}{\emph{Random-effects arm} ($n = 69$)} \\
\texttt{ols\_r}            & $1.00$ & $+0.08$ & $0.41$ \\
\texttt{re\_rho}$^\ddagger$& $1.00$ & $-0.14$ & $0.08$ \\
\texttt{tc\_rho}$^\dagger$ & $0.89$ & $+0.47$ & $0.51$ \\
\texttt{share\_significant}& $0.55$ & $+0.23$ & $2.10$ \\
\addlinespace
\multicolumn{4}{@{}l}{\emph{Tail-correlation arm} ($n = 63$)} \\
\texttt{ols\_r}            & $1.00$ & $-0.28$ & $0.19$ \\
\texttt{re\_rho}           & $0.87$ & $-0.93$ & $1.36$ \\
\texttt{tc\_rho}$^\ddagger$& $0.98$ & $+0.33$ & $0.31$ \\
\texttt{share\_significant}& $0.52$ & $+0.02$ & $2.05$ \\
\bottomrule
\end{tabular}
\end{table}

The $100\%$ coverage figures in Table~\ref{tab:calib} do not indicate good
calibration. Coverage above nominal means the simulated
interval is wider than the realized departures, and with $69$ candidates drawn
against the same $262$ experiments and the same north-star effects,
coverage is close to an all-or-nothing statistic: the departures share a
common component, so they tend to fall inside together or outside together.
That is why we report $\bar z$ and $\mathrm{sd}(z)$ instead of relying on
coverage. Read this way, the audit is informative. For
\texttt{ols\_r}, the baseline against which every correction is judged, the
archive's realized common departure is small in both arms ($+0.08$ and $-0.28$
simulated standard deviations), so the generator shows no systematic error,
and the idiosyncratic spread is well under one ($0.41$ and $0.19$), so no
individual candidate is badly reproduced.

There are two readings of $\mathrm{sd}(z) < 1$, and one archive cannot
separate them. The benign one is the common-shock decomposition above:
$\mathrm{sd}(z)$ estimates only the idiosyncratic part, so a value below one is
expected rather than diagnostic. The unfavorable one is that the generator
injects more replicate-to-replicate noise than the archive carries. If so, the
error has a known direction, and it flatters our results: a generator that is too
noisy makes every simulated ranking accuracy pessimistic and, because extra
noise acts like a smaller archive, pushes the crossover in
Figure~\ref{fig:crossover} toward larger archives than the true one,
overstating how far below it our archive sits. This is one reason we test the
simulated ordering on held-out real experiments
(Section~\ref{sec:horserace}), a test that does not use the generator.

The one clear calibration failure is the share of experiments in which a
candidate reaches significance: its coverage is only $52$--$55\%$, with
$\mathrm{sd}(z) \approx 2.1$ in both arms.
The failure lies in the spread across candidates, not in the average, which
is close ($\bar z = +0.23$ and $+0.02$): the generator's heavy-tailed
component has a single width, set to reproduce the share of significant
experiments averaged over candidates (Appendix~\ref{app:dgp}). Matching each
candidate's share would require candidate-specific widths, at the cost of the
exact true correlations against which every accuracy figure is scored. This
failure does not affect our conclusions: we use
\texttt{share\_significant} only on the real archive, as a reference statistic
that measures sensitivity rather than alignment, and never rely on its
simulated behavior.

\section{The domain and temporal transfer tests}
\label{app:domain}

This appendix reports the two transfer tests of Section~\ref{sec:horserace}:
across the archive's two product domains, and over time.

\paragraph{Transfer across product domains.} Ranked on the $78$ experiments of
the smaller domain and scored on the $184$ of the larger, every alignment
measure earns a positive reward, and the coarse ordering of the headline
comparison reappears: the uncorrected and observational measures lead, and
every estimator that removes part or all of $\Om$ trails them. The ordering
within these groups does not reappear: \texttt{ols\_r} moves ahead of both
observational measures, and the estimators that remove $\Om$ are no longer
ordered by dose. Because the domain split is deterministic, this is a single
comparison with no spread across splits to judge it by, so we read it as corroborating the
coarse ordering but not the finer one.

\paragraph{The reverse and temporal tests.} Both remaining tests are starved
of data on one side of the split. The reverse domain direction scores
candidates on only the $78$ experiments of the smaller domain, and its rewards
stay small and unstable however large the ranking set. The temporal test,
which ranks candidates on the experiments that ended before a cutoff and
scores them on those that followed, produces scores only for ranking sets of
$20$ or $50$ experiments, because on larger ones \texttt{tc\_rho} fails for
nearly every candidate (Section~\ref{sec:tcrho}). In both,
\texttt{share\_significant}, which carries no alignment content, earns a
positive reward, against a negative one in the headline comparison and in the
forward domain direction, so their rewards partly reflect something other than
alignment. We therefore draw no conclusions from either test, beyond noting
that \texttt{pda} and \texttt{tc\_rho} stay at or near the bottom of the
ordering in both. Random splits keep both sides large, which is why the
headline comparison uses them.

\end{document}